\documentclass[useAMS,usenatbib]{mn2e}

\def\kms{\hbox{km$\,$s$^{-1}$}}
\def\cm3{\hbox{cm$^{-3}$}}
\def\Apix{\hbox{\AA$\,$pix$^{-1}$}}

\newcommand\fsec{\hbox{$.\!\!^{\rm s}$}}
\def\one{\,{\sc i}}             
\def\two{\,{\sc ii}}
\def\three{\,{\sc iii}}

\usepackage{amsmath}
\usepackage{amssymb}
\usepackage{mathrsfs}
\usepackage{color}
\usepackage{upgreek}
\usepackage{graphicx}

\defcitealias{westm07a}{Paper~I}
\title[Mapping the galactic wind roots in NGC 1569]{Gemini GMOS/IFU spectroscopy of NGC 1569 -- II: Mapping the roots of the galactic outflow}
\author[M.S. Westmoquette et al.] {M.S. Westmoquette$^1$\thanks{E-mail: msw@star.ucl.ac.uk}, L. J. Smith$^{1,2}$ and J. S. Gallagher III$^3$, K.M. Exter$^{4,}$\thanks{Current address: Space Telescope Science Institute, 3700 San Martin Drive, Baltimore, MD 21218, USA}\\
$^1$Department of Physics and Astronomy, University College London, Gower Street, London, WC1E 6BT\\
$^2$Space Telescope Science Institute and European Space Agency, 3700 San Martin Drive, Baltimore, MD 21218, USA\\
$^3$Department of Astronomy, University of Wisconsin-Madison, 5534 Sterling, 475 North Charter St., Madison WI 53706, USA\\
$^4$Instituto de Astrof\'isca de Canarias, C/Via Lactea s/n, E38200 - La Laguna (Tenerife), Espa\~ na\\
}

\date{Accepted. Received; in original form}
\pagerange{\pageref{firstpage}--\pageref{lastpage}}
\pubyear{2007}
\begin{document}
\maketitle
\label{firstpage}
\begin{abstract}
We present a set of four Gemini-North GMOS/IFU observations of the central disturbed regions of the dwarf irregular starburst galaxy NGC 1569, surrounding the well-known super star clusters A and B. This continues on directly from a companion paper \nocite{westm07a}(Westmoquette et al.\ 2007a), in which we describe the data reduction and analysis techniques employed and present the analysis of one of the IFU pointings. By decomposing the emission line profiles across the IFU fields, we map out the properties of each individual component identified and identify a number of relationships and correlations that allow us to investigate in detail the state of the ionized ISM. Our observations support and expand on the main findings from the analysis of the first IFU position, where we conclude that a broad ($\lesssim$400~\kms) component underlying the bright nebular emission lines is produced in a turbulent mixing layer on the surface of cool gas knots, set up by the impact of the fast-flowing cluster winds. We discuss the kinematic, electron density and excitation maps of each region in detail and compare our results to previous studies. Our analysis reveals a very complex environment with many overlapping and superimposed components, including dissolving gas knots, rapidly expanding shocked shells and embedded ionizing sources, but no evidence for organised bulk motions. We conclude that the four IFU positions presented here lie well within the starburst region where energy is injected, and, from the lack of substantial ordered gas flows, within the quasi-hydrostatic zone of the wind interior to the sonic point. The net outflow occurs at radii beyond 100--200~pc, but our data imply that mass-loading of the hot ISM is active even at the roots of the wind. 
\end{abstract}


\section{Introduction}\label{intro}

Understanding feedback effects between massive stars, star clusters and the interstellar medium (ISM) is fundamental to the study of galaxy evolution. This is the second paper presenting integral field unit (IFU) observations of ionized gas in the dwarf starburst galaxy NGC 1569. Here we focus on the state of the ISM in the centre of the galaxy, by studying in detail four regions showing characteristics of wind--clump interaction.

The dwarf galaxy NGC 1569, is known to have undergone a significant recent global starburst event of $\gtrsim$100~Myr duration, the peak of which ended 5--10~Myr ago \citep{greggio98}. During this event the galaxy witnessed at least $10^4$ supernovae (SN) explosions, and the formation of many star clusters \citep{anders04}, including the two well-known super-star clusters (SSCs) A and B \citep{ables71, arp85}. The result of the energy input into the ambient medium can clearly be seen: NGC 1569 has a highly turbulent ISM, exhibiting ionized bubbles, arcs and filaments \citep{hunter93, heckman95}, and exhibits structure down to the resolution limit of every observation made so far.

\textit{HST} images, embodying the highest resolution observations currently available, illustrate well the complexity of the small-scale ISM in NGC 1569. \citet{buckalew06} used \textit{HST}/WFPC2 narrow-band imaging to examine the detailed structure of the ISM and the underlying ionizing mechanisms. They used the ratios of [S\two]/H$\alpha$ and [O\three]/H$\beta$ and the theoretical ``maximum starburst line'' \citep{kewley01} to determine a threshold between photoionized and non-photoionized (assumed to be shock excited) emission \citep[see also][]{calzetti04}. Identifying pixels in this way, they found structures of non-photoioinized points across the face of the whole galaxy. As well as finding a great deal of activity surrounding H\two\ region No.\ 2 \citep{waller91}, they found ``walls'' or arcs of non-photoionized points near SSCs A and B, indicating interaction between their respective cluster winds and the surrounding ISM. Another cluster of points was found between SSC A and cluster 10, indicating a further 
possible wind--wind interaction between these two clusters.

The distribution of He\two\,$\lambda$4686 emission in NGC 1569 was reported in an earlier contribution by \citet{buckalew00}, where its signature was used to search for Wolf-Rayet stars. Since He\two\ emission can also be produced in fast shocks \citep{garnett91}, \citet{buckalew06} conclude that where non-photoionized points were found to be coincident with He\two\ emission, this is strongly suggestive of shock-ionized gas. In a starburst environment, shock-excited emission dominates when the level of photoionization from very young stars starts to decline, and the energy input from stellar winds and SNe (which drive the shocks) is constant. This corresponds to the period 6--40~Myr \citep{leitherer99} and is consistent with the post-starburst classification and the typical ages of star clusters in NGC 1569 \citep{g-d97, anders04, angeretti05}. However, pockets of on-going star formation must still persist in smaller areas of the galaxy (e.g.~H\two\ region 2) to provide the continued photoionized component. This situation was confirmed by \citet{tokura06}, who, using observations made with the Subaru telescope, detected a number of strong mid-IR sources in NGC 1569, the most prominent of which (MIR1) is coincident with H\two{} region 2. This and other detections indicate continued, deeply embedded star formation in H\two{} region 2 and the surrounding regions.

As well as expanding bubbles and shells of shocked gas, evidence of the past starburst activity can also be found through other, more direct ways. \citet{greve02} identify a number of radio supernovae (SNe) and supernova remnants (SNRs) distributed throughout the NGC 1569 disc using MERLIN and VLA 1.4~GHz and 5~GHz observations. The number of non-thermal sources identified throughout the central regions indicate that a great deal of star formation has taken place in the SSC A/B region in the past.

The collective effect of the massive stars in the clusters formed in the starburst has injected a great deal of kinetic energy and momentum into the surrounding gas, together with chemically processed material and copious ionising radiation. This has resulted in an impressive galaxy-wide outflow seen to extend over kpc scales in both H$\alpha$ \citep{martin98} and X-rays \citep{heckman95, martin02}.

\citet{heckman95} also presented deep H$\alpha$ long-slit spectroscopy of the disc and filament system, and found the H$\alpha$ line-profiles within the galaxy core to be ``distinctly non-Gaussian''. Decomposing them using multi-Gaussian profile fitting allowed them to identify two kinematically distinct subsystems: a bright and centrally located `quiescent system' centred on SSC A, with line widths of FWHM = 30--90~\kms\ and radial velocities within 30~\kms\ of their adopted $v_{\rm sys}$ ($-68$~\kms); and a fainter and extended `high-velocity system' with FWHMs $\leq 150$~\kms\ and velocities up to 200~\kms\ relative to their $v_{\rm sys}$. This confirms the finding of \citet{tomita94} who also report two similar velocity systems. Furthermore, at the location of SSC A, they found the H$\alpha$ profile to have weak but very broad wings. The investigation of this enigmatic, underlying broad component formed one of the main focusses of the first part of this study described in \nocite{westm07a}Westmoquette et al.\ (2007a, hereafter \citetalias{westm07a}).

In \citetalias{westm07a}, we focussed on one IFU position, covering the area surrounding cluster 10 just to the east of H\two\ region 2. Through accurately modelling the observed line profiles we were able to map the emission line properties of each line component across the IFU field-of-view (FoV). We found that the observed nebular emission lines are composed of a bright, narrow component and an underlying, very broad ($\lesssim$300~\kms) component. By mapping out, for the first time, the properties of both the narrow component (which we refer to as C1) and the broad component (referred to as C2) in two dimensions, we were able to perform a detailed investigation of each component and their relationship to one another, and develop a new semi-unified model of the state of the ionized gas clumps in the ISM to explain the origin of the line profiles. We concluded that the bright, narrow component (C1) represents the ambient turbulent ISM, where its motions (line width) result from a convolution of the general stirring effects of the starburst and gravitational virial motions. We found that the width of C2 is both highly correlated with the intensity of C1 and anti-correlated with its own flux. This led us to conclude that the most likely explanation for C2 is that it originates from a turbulent mixing layer \citep{begelman90, slavin93, binette99} set up on the surface of the gas clumps by the effects of the strong cluster winds. In this interpretation, the turbulent gas is then stripped from this interface layer through hydrodynamical effects such as thermal evaporation and mechanical ablation as it becomes entrained in the wind flow, resulting in a diffuse, turbulent gas component pervading the whole region.

In this paper, we extend our investigation to the remaining three IFU fields. In Section~\ref{sect:data} we describe the observations taken and summarise the data-reduction process described in detail in \citetalias{westm07a}. Then, in Section~\ref{sect:results}, we perform a detailed study of each remaining IFU position and, armed with our findings from \citetalias{westm07a}, investigate how our prior conclusions apply to the other regions within the starburst core.

Previous studies of the ISM in starburst galaxies and their resulting outflows have tended to concentrate on large-scale features (mainly due to instrumental restrictions). Hence there exists a paucity of data of sufficient quality and resolution to compare directly to the latest generation of high-resolution hydrodynamic starburst-driven outflow models \citep[e.g.][]{stevens03,t-t03}. By examining the collective state of the ISM as shown by all four fields together in Section~\ref{sect:ISM_state}, we attempt to begin filling in these observational deficiencies. Finally our results and conclusions are summarised in Section~\ref{sect:summary}.

In a third paper (Westmoquette et al. in prep.; Paper III) we will examine the properties of the galactic outflow as a whole using deep H$\alpha$ imaging and IFU observations and attempt to create a unified physical picture of the outflow state by connecting what we find here in the central regions to the outer halo.

\begin{figure*}
\centering
\includegraphics[width=17cm]{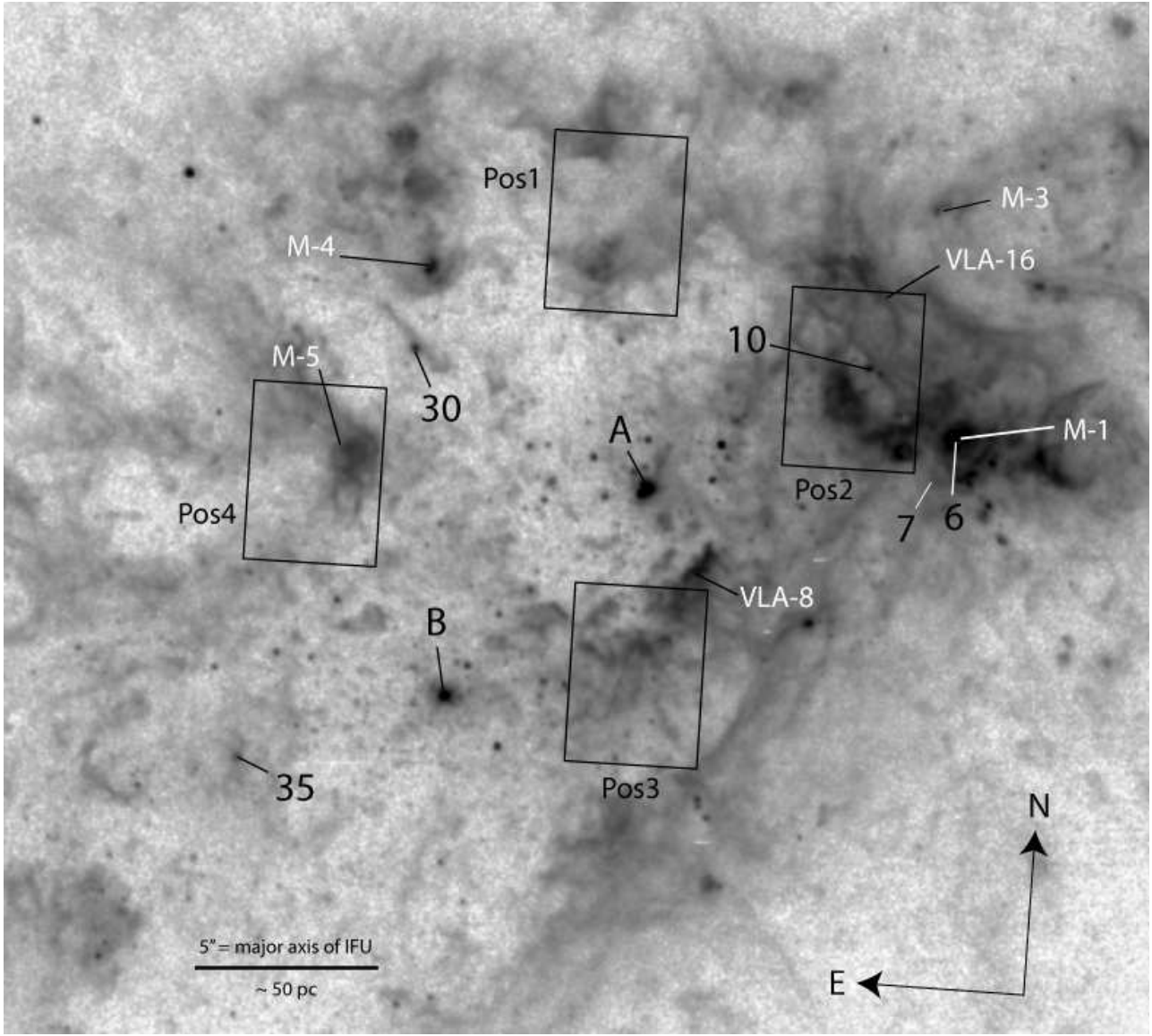}
\caption{\textit{HST}/WFPC2 F656N image of the centre of NGC 1569 showing the positions of the four IFU fields. A number of the most prominent star clusters \citep{hunter00} and radio continuum sources \citep{greve02} have been labelled in black and white, respectively}
\label{fig:GMOS_finder}
\end{figure*}

\section{Observations and Data Reduction} \label{sect:data}

In October and November 2004 we obtained Gemini-North Multi-Object Spectrograph (GMOS) data covering four regions in the nucleus of NGC 1569 with the integral field unit \citep[IFU; ][]{allington02} (programme ID: GN-2004B-Q-33, PI: L.J.\ Smith). A full description of the instrument and the set-up is given in \citetalias{westm07a}, but briefly, the IFU in one-slit mode gives a field-of-view of $5\times 3.5$~arcsecs (which corresponds to approximately \mbox{$50\times 35$~pc} at the distance of NGC 1569) sampled by 500 hexagonal contiguous fibres of $0\farcs2$ diameter. A separate offset block of sky fibres provides a simultaneous sky view. We took four exposures at each position with integration times between 940 and 1680s each, and used the R831 grating to give a spectral coverage of 4740--6860~\AA\ and a dispersion of 0.34~\Apix. Table~\ref{tbl:gmos_obs} lists the coordinates and exposure times for each position, and their respective locations are shown in Fig.~\ref{fig:GMOS_finder}.

Position 1 was chosen to intercept the (possibly) high velocity outflow emanating from the central regions of the galaxy. Position 2 is coincident with cluster 10 \citep{hunter00} and covers some of the H$\alpha$ gas on the easternmost part of the ionized complex centred on H\two\ region 2 \citep{waller91}. A detailed description and analysis of this position, including a discussion of the relationship between cluster 10 and its environment, is given in \citetalias{westm07a}. Position 3 was placed to the south of SSC A, and covers a number of compact ionized knots. VLA-8, a non-thermal radio source which \citet{greve02} attribute to being low surface brightness SNR, is located at the very top of the field-of-view. IFU position 4 was chosen to focus on an ionized knot to the east of SSC A, north of SSC B, and south of cluster 30. We suspected this region might show signs of the interaction of winds from these three star clusters. This knot also contains the non-thermal radio source M-5 \citep{greve02}.

\begin{table}
\centering
\caption {Gemini GMOS observations}
\label{tbl:gmos_obs}
\begin{tabular}{c c r @{\hspace{0.2cm}} l r @{ $\times$ } l}
\hline
Pos. & Date & \multicolumn{2}{c}{Coordinates} & \multicolumn{2}{c}{Exp Time} \\
No. & & \multicolumn{2}{c}{(J2000)} & \multicolumn{2}{c}{(s)} \\
\hline 
1 & 7/10/04 & $04^{h}30^{m}48\fsec38$ & $64^{\circ}51'06\farcs5$ & 1600 & 4 \\
2 & 7/11/04 & $04^{h}30^{m}47\fsec27$ & $64^{\circ}51'03\farcs2$ & 1680 & 4 \\
3 & 7/10/04 & $04^{h}30^{m}48\fsec04$ & $64^{\circ}50'54\farcs8$ & 940 & 4 \\
4 & 8/11/04 & $04^{h}30^{m}49\fsec58$ & $64^{\circ}50'58\farcs9$ & 1680 & 4 \\
\hline
\end{tabular}
\end{table}

A full description of the data reduction steps is given in \citetalias{westm07a}. Here we briefly summarise the procedure and describe the changes in the differential atmospheric refraction (DAR) correction we had to make for the remaining fields. Basic reduction was performed following the standard Gemini reduction pipeline (implemented in \textsc{iraf}\footnote{The Image Reduction and Analysis Facility ({\sc iraf}) is distributed by the National Optical Astronomy Observatories which is operated by the Association of Universities for Research in Astronomy, Inc. under cooperative agreement with the National Science Foundation.}). First, a trace of the position of each spectrum on the CCD was produced from the flat field. Throughput correction functions and wavelength calibration solutions were created and applied to the science data, before the individual spectra were extracted to produce a data file containing 750 reduced spectra. The final steps were to clean the cosmic-rays, subtract an averaged sky spectrum (computed from the separate sky field), and apply a flux calibration derived from observations of the standard star G191-B2B. Separation of the sky spectra from the science data resulted in a data file formed of 500 fully reduced spectra, with a resolution varying between FWHM = 74\,$\pm$\,5~\kms\ at the blue end (with an average S/N of 4.6 in the continuum), to 59\,$\pm$\,2~\kms\ at the red end (with an average S/N of 14.5).

DAR is the effect of varying light refraction by the Earth's atmosphere at different wavelengths. In order to correct our data for DAR, we had to apply a spatial interpolation to resample the data from the hexagonal fibre arrangement into contiguous squares. The methods are fully described in \citetalias{westm07a}, together with how the DAR shift was measured in position 2 by tracing the position of cluster 10 through wavelength space to produce a DAR curve. Unfortunately none of the other fields contain an unresolved source suitable for tracing the effects of DAR, so we had to employ a visual inspection method to determine the DAR shift. The Euro3D Visualisation Tool \citep{sanchez04} allows a spatial intensity map of a specific wavelength range for each IFU field to be plotted. After selecting two maps of the field isolating the H$\beta$ and H$\alpha$ emission lines respectively, we measured the offset in position of a representative bright gas knot between these two known wavelengths. Assuming that these measurements represent the wavelength extrema (a reasonable approximation for our set-up), we normalised the curve output from the automatic DAR tracking algorithm to our measured values in order to get the DAR corrections for these positions. The calculated offsets from H$\beta$ to H$\alpha$ for each field are given in Table~\ref{DARcor}.

\begin{table}
\centering
\caption {Differential atmospheric refraction (DAR) corrections (H$\beta$ to H$\alpha$). 1 spaxel has a diameter of $0\farcs18$.}
\label{DARcor}
\begin{tabular}{c r @{, } l}
\hline
Position & \multicolumn{2}{c}{Spaxel Shift} \\
\hline 
1 & 1.50 east & 2.50 north \\
2 & 0.39 west & 2.98 north$^{\dagger}$ \\
3 & 0.50 west & 2.50 north \\
4 & 1.15 west & 2.75 north \\
\hline
\end{tabular}

$^{\dagger}$ measured using the centroid fitting routine described in \citetalias{westm07a}.
\end{table}

\section{Properties of the ionized gas} \label{sect:results}

\subsection{Decomposing the line profiles} \label{sect:profile}
In order to quantify the gas properties observed in each IFU field, we fitted multiple Gaussian profile models to each emission line using a program written in IDL called \textsc{pan} \citep[Peak ANalysis;][]{dimeo}. A detailed description of the program and our modifications is given in \citetalias{westm07a}, together with the statistical techniques we employed to determine the optimum number of Gaussian components to fit to each line.

The general shape of the emission line profiles observed in all four positions is a convolution of a bright, narrow component overlying a faint, broad component. In \citetalias{westm07a}, we describe our methodology and nomenclature for defining initial guess parameters for single-, double- and triple-Gaussian fits. To summarise, each line in each of the 500 spectra was fitted using a single-, double- and triple-Gaussian component initial guess, where line fluxes were constrained to be positive and widths to be greater than the instrumental contribution. When a double component fit was deemed most appropriate, fits were always specified with the first Gaussian as the narrow component (which we refer to as C1), and the second Gaussian as a broader component (referred to as C2). Never did we find a double-peaked broad line. For triple component fits, an additional component (C3) was specified with a initial guess assigning it to a supplemental narrow line at the same wavelength as the main narrow line (C1). We employed the statistical F-test, giving the minimal increase of the fit $\chi^{2}$ ratio that would be required for one fit to be statistically distinguishable from another, together with a number of additional physical tests, to determine the best number of components to fit to each line. Clearly the quality of the fits depends significantly on the S/N of the emission line. We estimate the uncertainties on the fit properties in \citetalias{westm07a} (section~3.1); to summarise, we found that for the H$\alpha$ line flux, uncertainties vary between 0.5--10 per cent for C1 (for high--low S/N lines), 8--15 per cent for C2 and 10--80 per cent for C3. FWHM errors vary between 0.25--2~\kms\ for high--low S/N C1 lines respectively, 4--15~\kms\ for C2 and 15--20~\kms\ for C3. For the radial velocity, the error in C1 varies between 0.1--3~\kms, 2--5~\kms\ for C2 and 10--30~\kms\ for C3. We estimated that the addition of a third component (where required) increases the errors in the C1 and C2 flux by 5--10 per cent, FWHM by $\sim$5~\kms\ and the central velocity by $\sim$3~\kms.

Once the profile of each line in each spectrum was decomposed using this method, we used the visualisation tool, \textsc{daisy}, also described in \citetalias{westm07a}, to map the spatial variation of the Gaussian fit properties and derived values for each GMOS position. The flux, width and velocity maps of all the emission lines look qualitatively similar, so we concentrate primarily on the H$\alpha$ line since it has the highest S/N (see also \citetalias{westm07a}). In the following sections, we discuss the observational characteristics of IFU positions 1, 3 and 4 in turn with reference to the aforementioned maps. In each case, we begin by describing the results, including any trends and correlations seen, then attempt to interpret what we find and describe how it relates to the central region as a whole.

Following \citetalias{westm07a}, we adopt a systemic velocity, $v_{\rm sys}=-80$~\kms\ throughout the analysis.

\subsection{GMOS Position 1} \label{sect:GMOS_pos1}

\begin{figure*}
\includegraphics[width=16cm]{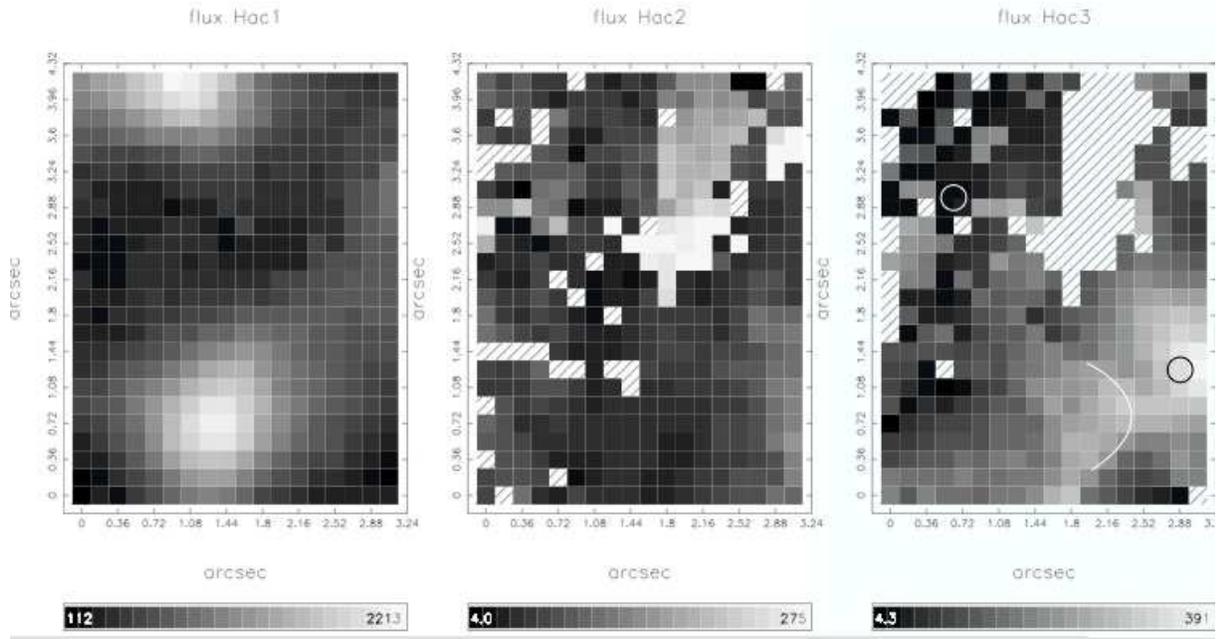}
\caption{{\bf Pos 1.} \emph{Left:} Flux map in H$\alpha$ C1 (range 112--2213); \emph{centre:} flux map in H$\alpha$ C2 (range 4--275); \emph{right:} flux map in H$\alpha$ C3 (range 4.3--391). Non-detections are represented as hatched spaxels, the $x$ and $y$ scales are in arcseconds offset from the lower-left spaxel, and a scale bar is given for each plot with units $10^{-15}$~erg~s$^{-1}$~cm$^{-2}$~spaxel$^{-1}$. North is up and east is left. The spaxels from which the example H$\alpha$ profiles shown in Fig.~\ref{fig:1fit_egs} were extracted are marked with circles, and the semi-circular feature discussed in the text is outlined with a white line.}
\label{fig:1ha_flux}
\end{figure*}

\begin{figure*}
\includegraphics[width=16cm]{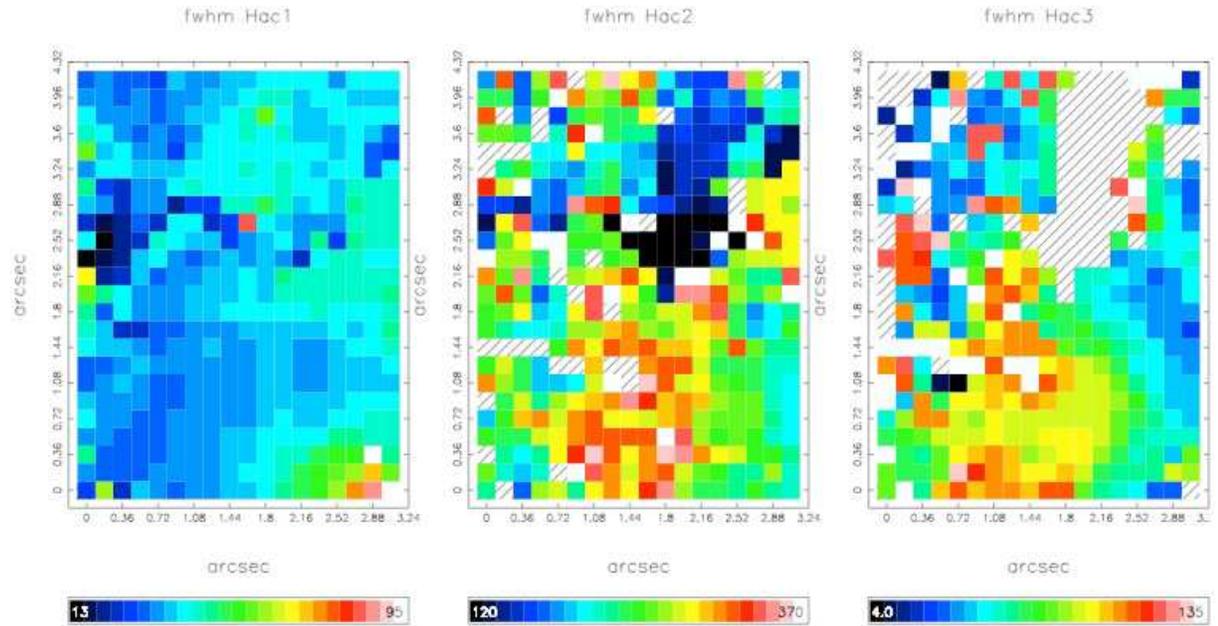}
\caption{{\bf Pos 1.} \emph{Left:} FWHM map in H$\alpha$ C1 (range 13--95); \emph{centre:} FWHM map in H$\alpha$ C2 (range 120--370); \emph{right:} FWHM map in H$\alpha$ C3 (range 4--135). A scale bar is given for each plot in \kms, and the measurements are corrected for instrumental broadening. North is up and east is left.}
\label{fig:1ha_fwhm}
\end{figure*}

\begin{figure*}
\includegraphics[width=16cm]{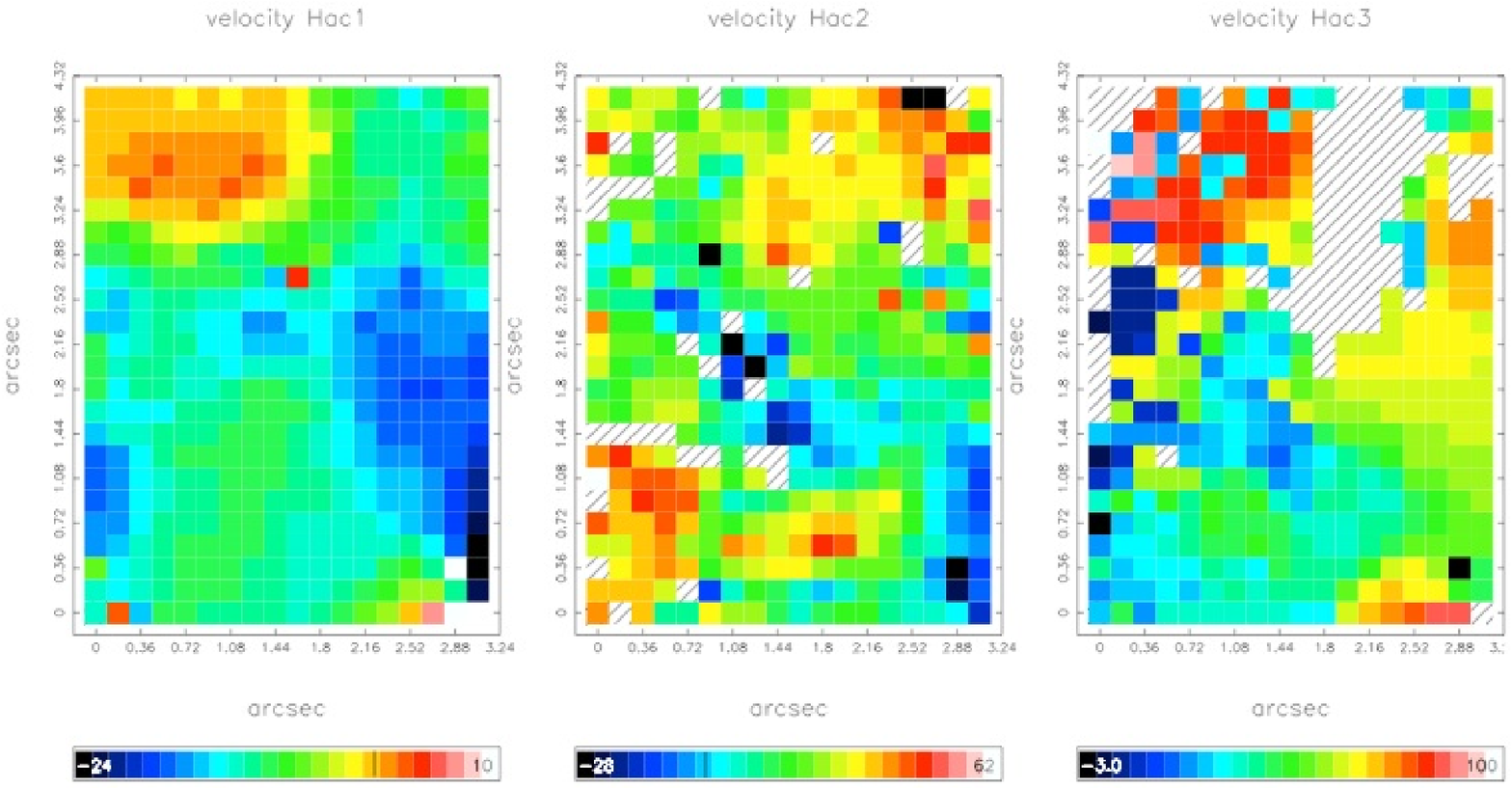}
\caption{{\bf Pos 1.} \emph{Left:} radial velocity map in H$\alpha$ C1 (range $-24$--10); \emph{centre:} radial velocity map in H$\alpha$ C2 (range $-28$--62); \emph{right:} radial velocity map in H$\alpha$ C3 (range $-3$--100). A scale bar is given for each plot in \kms{} (heliocentric, relative to the systemic velocity of the galaxy, $v_{\rm sys} = -80$~\kms), where zero is marked with a line. North is up and east is left.}
\label{fig:1ha_vel}
\end{figure*}

Position 1, the most northerly of our four positions, coincides with two large knots of gas. The `swept-back' morphology of these (and other nearby) knots suggests that they have been shaped by interactions with the material being evacuated from the region surrounding SSC A, and this position was specifically chosen to look for evidence of these interactions. Figs~\ref{fig:1ha_flux}--\ref{fig:1ha_vel} show maps of the Gaussian properties (flux, FWHM and radial velocity) for each line component over the whole FoV. The two gas knots show up clearly in the C1 flux map (Fig.~\ref{fig:1ha_flux}, left panel), and provide convenient reference points for interpretation of the other maps. Unlike the other positions, we require not only a bright narrow component (C1) and a broad underlying component (C2) to fit the H$\alpha$ line profile, but a third Gaussian was needed to fit the integrated line profile over a significant proportion of the field. Because of the additional uncertainties introduced by fitting a third component, we precede our discussion of the results with a brief description of how this has affected the emission-line property maps and how this affects our interpretation of them.

The discontinuous region towards the central-north-west of the field seen in the C2 flux and FWHM maps is caused by the sudden change from fitting with triple- to double-Gaussian profiles. According to the F-test method \citepalias[see][]{westm07a}, the $\chi^{2}$ ratio in this region became no longer sufficient to justify a triple-Gaussian fit. The artificially high flux and low FWHM in C2 has resulted from fitting two components to an intrinsically three-component profile, since C2 now includes some contribution from the ignored third component. The continuous nature of the radial velocity maps (which are less affected by poor S/N) proves that the component assignments were correctly made over the whole field and that we can safely ignore only the discontinuous region in the maps in the following discussion.


\begin{figure*}
\begin{minipage}{6cm}
\includegraphics[width=6cm]{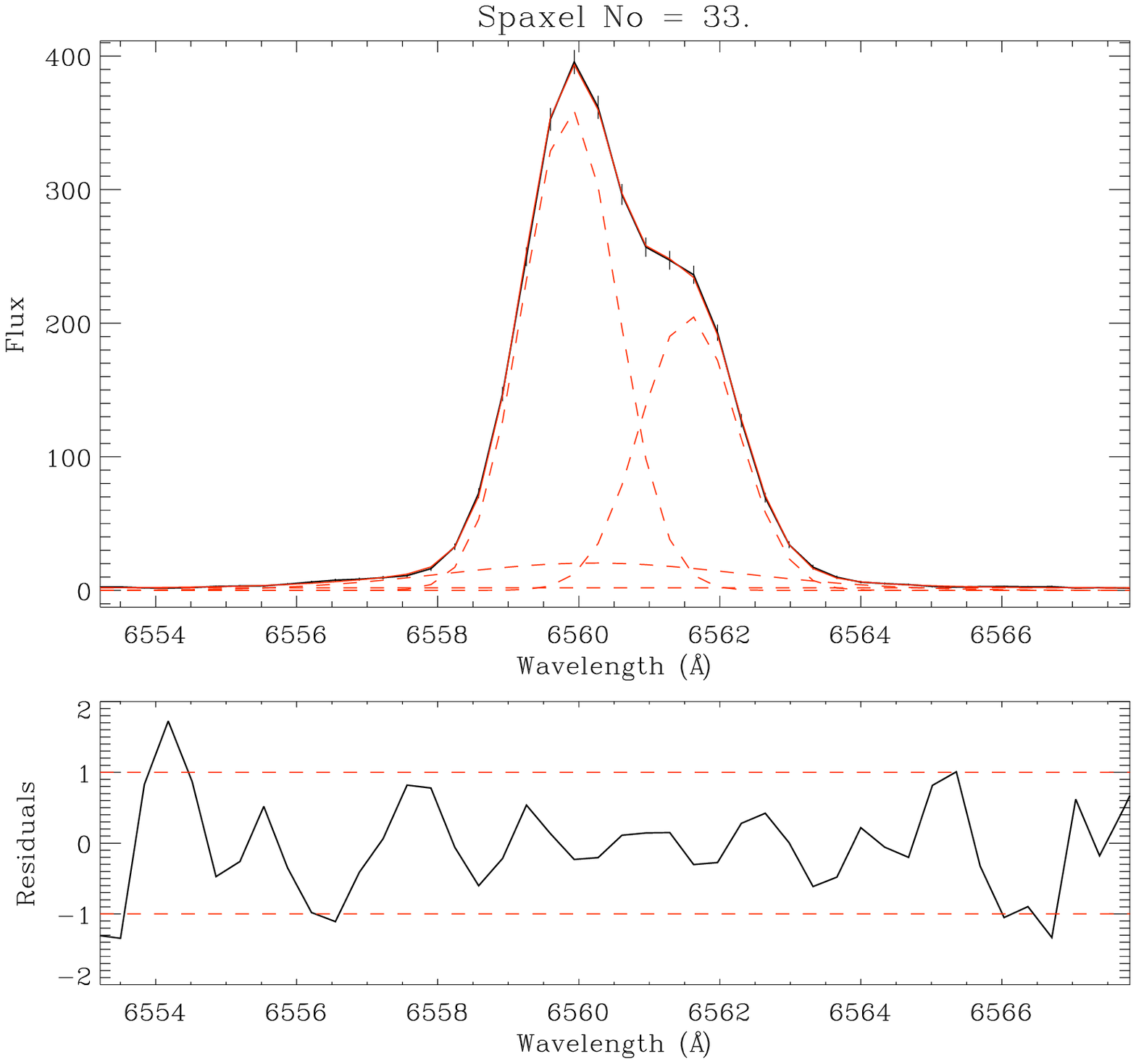}
\end{minipage}
\hspace*{0.3cm}
\begin{minipage}{6cm}
\includegraphics[width=6cm]{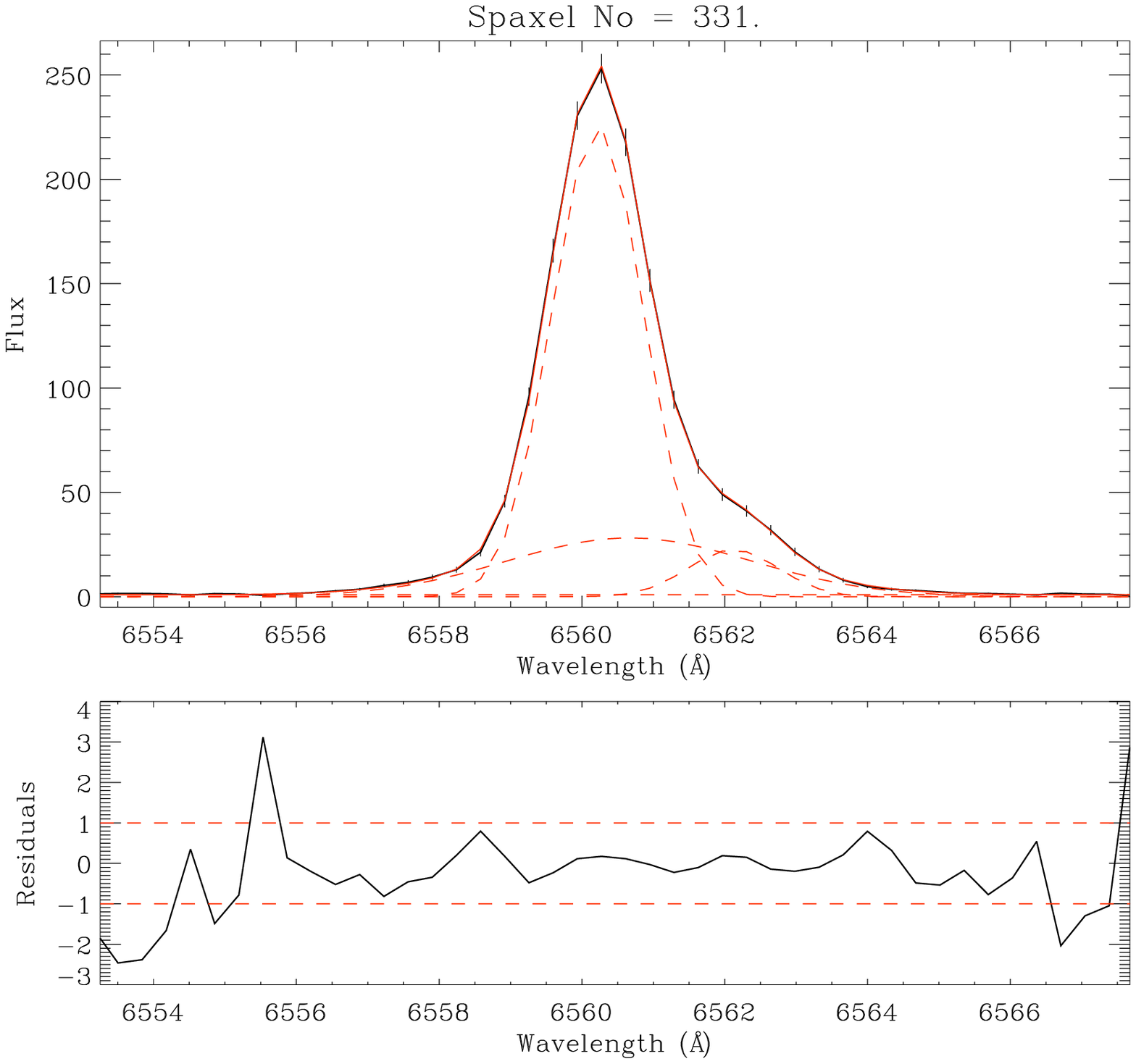}
\end{minipage}
\caption{{\bf Pos 1.} Examples of H$\alpha$ fits for spaxels 33 and 331 (positions marked with circles on Fig.~\ref{fig:1ha_flux}). These examples illustrate the presence of three significant line components and their contribution to the integrated line profile.}
\label{fig:1fit_egs}
\end{figure*}

The C1 flux map (Fig.~\ref{fig:1ha_flux}, left panel) highlights the two gas condensations, which, without any dynamical information appear very similar. From the equivalent FWHM (Fig.~\ref{fig:1ha_fwhm}, left panel) and radial velocity (Fig.~\ref{fig:1ha_vel}, left panel) maps, it is clear that the two do not share the same characteristics. The southern knot stands out for having the broadest (350--400~\kms) C2 widths of all the fields (see Fig.~\ref{fig:1ha_fwhm}, centre panel) and for having strong associated emission in C3. However the peak in C3 emission (Fig.~\ref{fig:1ha_flux}, right panel), found just to the west of this knot, does not have a counterpart in the \textit{HST} F656N image (Fig~\ref{fig:GMOS_finder}). Between this bright area and the C1 knot, a faint semi-circular structure can be identified (indicated on Fig.~\ref{fig:1ha_flux} with a white line), with a concentric morphology centred on the southern C1 knot. The FWHM of C3 increases radially towards the southern knot following this concentric pattern,
while its radial velocity decreases slightly, resulting in a velocity difference between C1 and C3 at the knot's location of 45--55~\kms. Fig.~\ref{fig:1fit_egs} shows two examples of H$\alpha$ line profiles from this position, including one from this bright C3 region (the location of the spaxels from which the profiles are from are marked with circles on the right-hand panel of Fig.~\ref{fig:1ha_flux}). These examples clearly show the existence of three components; the high quality of the fits can be qualitatively assessed through inspection of the residual plots shown below the main plot in each case (see \citetalias{westm07a} for an explanation of how the residuals are calculated). In the southern knot, there is a very good correlation between C1 flux and C2 FWHM (i.e.~the brightest C1 knots have the widest C2 FWHM), and anti-correlation between C2 flux and C2 FWHM (i.e.~the brightest regions in C2 have the narrowest C2 width), as we found for position 2 \citepalias{westm07a}.

The northern C1 knot has a redshifted velocity quite distinct from its surroundings (+5~\kms, compared to the field average of $\sim$$-15$~\kms), but the velocity of C3 at this point is even more redshifted ($\sim$+100~\kms). At the location of the knot, the C1--C3 velocity difference is $\sim$90~\kms{} (an example of a line profile from this region is shown in Fig.~\ref{fig:1fit_egs}, right panel). If this difference results from expansion, then C1 represents the near side of the northern clump, and it must be located further along the line-of-sight than the southern knot and be expanding with a velocity, $v_{\rm exp} \approx 45$~\kms{} (half the C1--C3 velocity difference).

The properties of the inter-clump gas are generally fairly uniform. The only unusual relationship is that C1 has a blueshifted ($\sim$$-15$~\kms) mean velocity, whereas C2 and C3 are both redshifted relative to $v_{\rm sys}$ \citepalias[-80~\kms][]{westm07a}. A region with distinctly different characteristics is found in the far south-west. Here the gas becomes rapidly redshifted in both C1 and C3, and is coincident with an area of very broad ($\sim$100~\kms) C1 emission. This corner is faint in all three line components, and by comparing its location to the \textit{HST} image, appears to be part of the low surface-brightness cavity surrounding SSC A.

\begin{figure*}
\begin{minipage}{5.5cm}
\includegraphics[width=5.5cm,height=8cm]{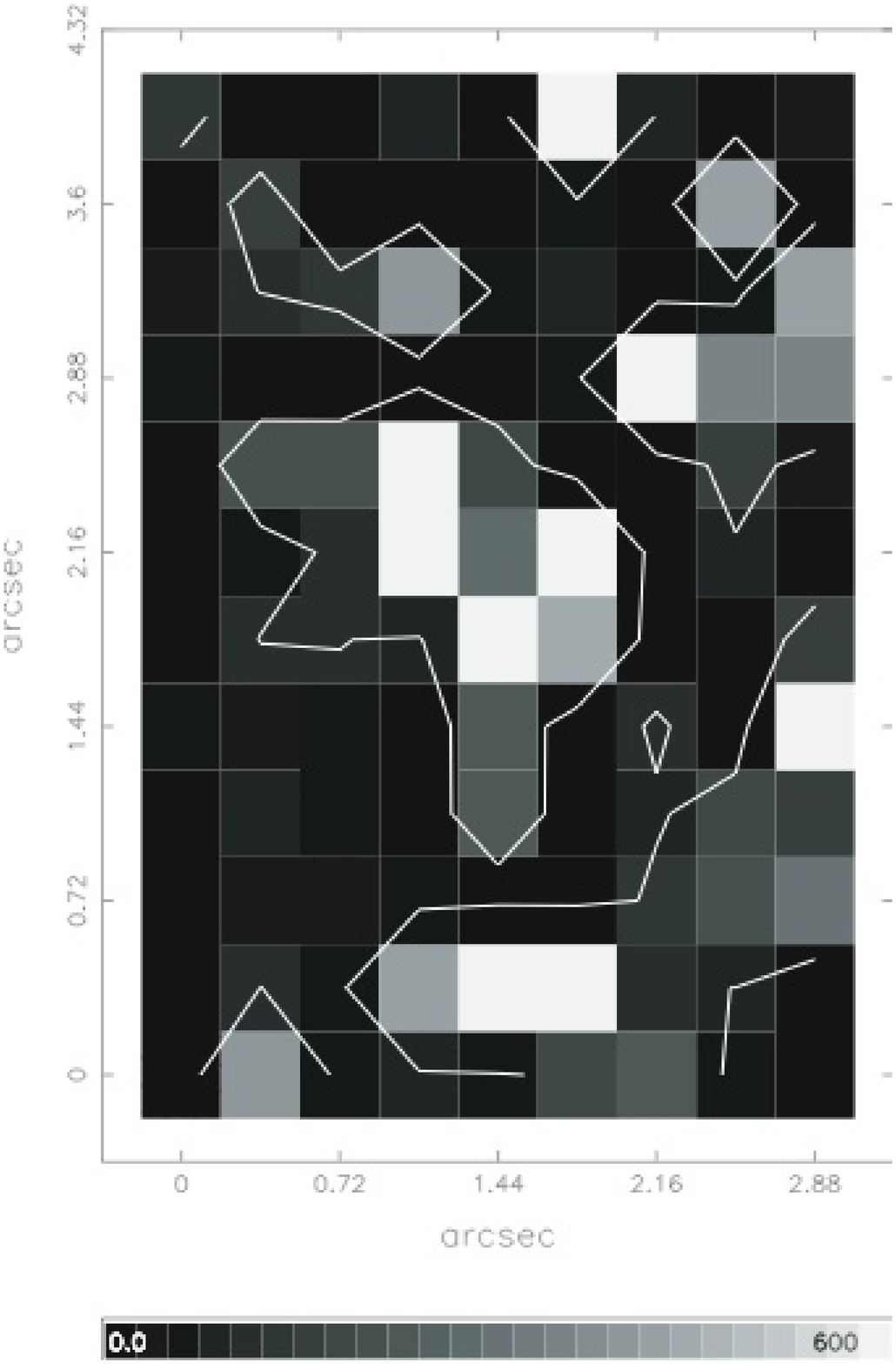}
\end{minipage}
\hspace*{0.8cm}
\begin{minipage}{5.5cm}
\includegraphics[width=5.5cm,height=8cm]{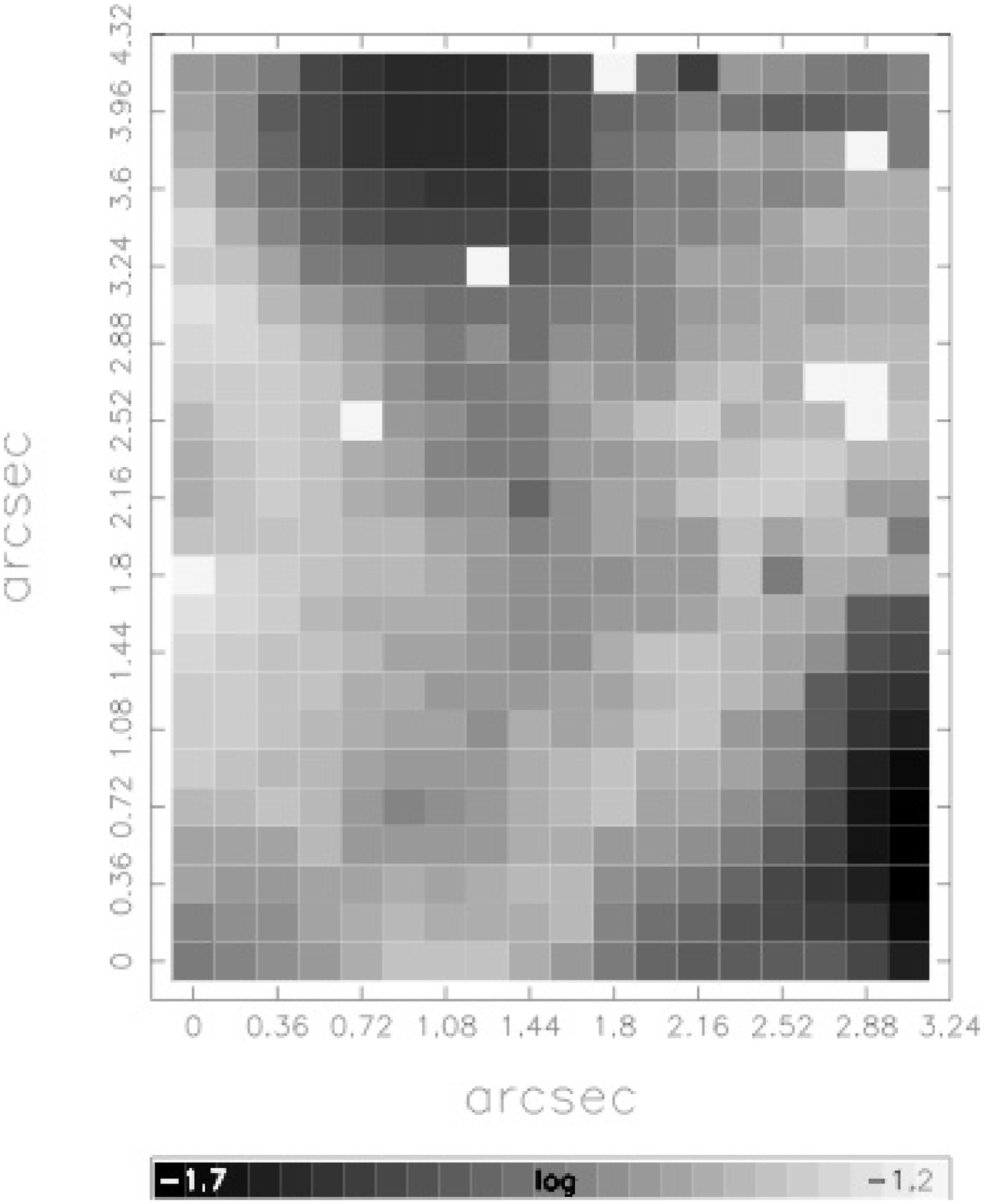}
\end{minipage}
\caption{{\bf Pos 1} \emph{Left:} Electron density map derived from the ratio of [S\two]$\lambda6717 / \lambda6731$ for C1 [a scale bar is given in units of cm$^{-3}$; a single contour line indicates the low density limit \citep[100~\cm3;][]{osterbrock89}; \emph{right:} ratio of [S\two]$\lambda$6717+$\lambda$6731/H$\alpha$ in logarithmic scale.}
\label{fig:1elecdens}
\end{figure*}

The electron density of the ionized gas can be derived from the flux ratio of [S\two]$\lambda$6713/$\lambda$6731, but the S/N of the [S\two] lines in this field were found to be too low to make any confident profile fits. We have therefore spatially binned the datacube by 2$\times$2 in order to increase the S/N per spaxel, and fitted the resultant [S\two] profiles with \textsc{pan} as we did for H$\alpha$. The resulting map for C1 only is shown in Fig.~\ref{fig:1elecdens} (left panel). The C1 electron densities across the field are in general below or at the low density limit (indicated by the single contour line), with the notable exception of three regions where the densities reach a few hundred \cm3. One is found in the south of the field extending up towards the north-west, coincident with the bright C1 and C3 emission associated with the southern knot; the other two are located in the centre and north-west of the field, apparently not associated with any bright emission at all. This result shows that the compact nature of the two knots seen in the flux map of C1 is in fact misleading, and that the density of the gas in these condensations (particularly the northern knot) is actually quite low.

Line flux ratios can also be used to trace the ionization level of the emitting gas. In Fig.~\ref{fig:1elecdens} we show the map of [S\two]$\lambda$6717+$\lambda$6731/H$\alpha$. Since not all individual components could be identified in all lines, we calculated the ratios from the sum of the flux in all observed line components. Furthermore, this diagnostic is less sensitive to low S/N since we sum rather than divide the flux from the two [S\two] lines, and we have plotted this map without having to bin the datacube. The lowest ratios are found towards the north-east of the field, corresponding to the northern H$\alpha$ knot, and in the south-west, roughly coincident with the peak in H$\alpha$ C3 emission. This ratio, when combined with [O\three]$\lambda$5007/H$\beta$ (not shown here because of the low S/N of the H$\beta$ line) can be used to look for indications of shocks. Following the method described in \citetalias{westm07a} for position 2, we define points in [O\three]/H$\beta$--[S\two]/H$\alpha$ space as containing a contribution from non-photoionization processes when they fall above the `maximum starburst line' \citep{kewley01}. All points in this field lie below this line. We noted in \citetalias{westm07a} that the low metallicity of NGC 1569 may mean that this threshold for non-photoionized (shocked) emission is in reality lower than the generalised `maximum starburst line'. We therefore conclude that emission in this field is close to being fully consistent with pure photoionization. A contribution due to shocks may be present, but is currently unquantifiable.

\subsubsection{Interpretation}
This position covers two dynamically different gas knots that form part of the clumped medium around which the central cavity is being evacuated. The southern half of this field is similar to position 2 in that we see very broad (up to 350~\kms) C2 widths coincident with bright C1 emission. In \citetalias{westm07a} we discussed in detail what possible mechanisms might cause such broad emission coincident with brightly emitting knots, and concluded that the most likely explanation is a turbulent mixing layer on the surface of the gas cloud set up by the shear forces created by the fast-flowing cluster winds. Thermal evaporation and/or mechanical ablation of clump material result in mass being stripped off the surface of the clouds and thus a highly turbulent velocity field pervades the whole area. We note that in every case (here and in position 2), the broad lines are single peaked, meaning we are not likely to be seeing line broadening due to projection through a bipolar outflow. Since the data for this southern knot are fully consistent with the turbulence explanation, we can extend our conclusions taking into account a number of other observations.

Firstly, in general the gas is found to have a low density, possibly explaining why the average intensity of C1 and C2 are respectively five and three times lower than in position 2. These low intensities have resulted in more uncertain fits and much noisier maps. A density enhancement of $\sim$500~\cm3\ is observed in the centre of the southern clump, but the gas immediately surrounding it is of very low density ($<$100~\cm3). Secondly, the width of the broad component over the whole field is on average 50~\kms{} larger than in position 2. We interpret these two facts to mean that this southern knot is undergoing a strong interaction with the cluster winds from the central region (particularly SSC A) and is at a more advanced stage in its dissolution than the clumps observed in position 2. The low density of the outer parts of the gas knot may mean that the wind can penetrate more of the clump and material is being stripped off more easily.

We do not see evidence for the swept-back tail morphology seen in the \textit{HST} H$\alpha$ image, but this may again be due to the very low density, diffuse nature of the gas or the limited spatial-resolution of our IFU observations. Furthermore, as shown by the random distribution of radial velocities measured in this region, we cannot be sure which cluster or clusters are dominating the energetics in this region, so it is quite possible that what might appear to be a swept-back, bow-shaped interaction in the \textit{HST} image is just part of the ambient, clumped gas distribution.

Observing such a strong additional narrow component (C3) over the majority of the field is an indicator of the disturbed nature of this part of the inner 200~pc, however we defer a discussion of this component until Section~\ref{sect:ISM_state} where we consider it in context with the whole central region.

\subsection{GMOS Position 3} \label{sect:GMOS_pos3}

Position 3 is located just to the south of SSC A, in what appears to be a region of turbulent filamentary gas. The radio source VLA-8 is located near the north of the field at coordinates $\alpha = 4^{\rm h}\,30^{\rm m}\,48\fsec0$, $\delta = 64^{\circ}\, 50'\,56\farcs3$ \citep[][see Fig.~\ref{fig:GMOS_finder}]{greve02}, and is thought to be an extended ($2''\equiv 20$~pc), low surface brightness SNR. To the south and west, just beyond the IFU FoV, there is a large, prominent, semi-circular shell-like structure outlined by a string of non-photoioinized points \citep{buckalew06}, indicating that this could be part of a fast, expanding bubble. Due to its close projected distance from SSC A, this region looks like it may bear the brunt of the impacting winds from the central star clusters.

\begin{figure*}
\includegraphics[width=11cm]{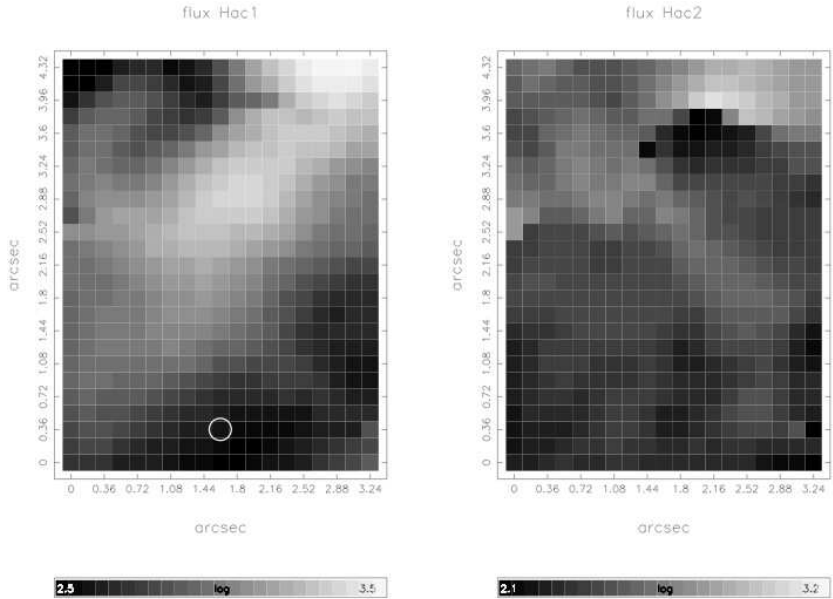}
\caption{{\bf Pos 3.} \emph{Left:} Flux map in H$\alpha$ C1 (log range 2.5--3.5); \emph{right:} flux map in H$\alpha$ C2 (log range 2.1--3.2). The $x$ and $y$ scales are in arcseconds offset from the lower-left spaxel, and a scale bar is given for each plot with units log $10^{-15}$ erg s$^{-1}$ cm$^{-2}$~spaxel$^{-1}$. North is up and east is left. The spaxel from which the example H$\alpha$ profile shown in Fig.~\ref{fig:3fit_egs} was extracted is marked with a circle.}
\label{fig:3ha_flux}
\end{figure*}

\begin{figure*}
\includegraphics[width=11cm]{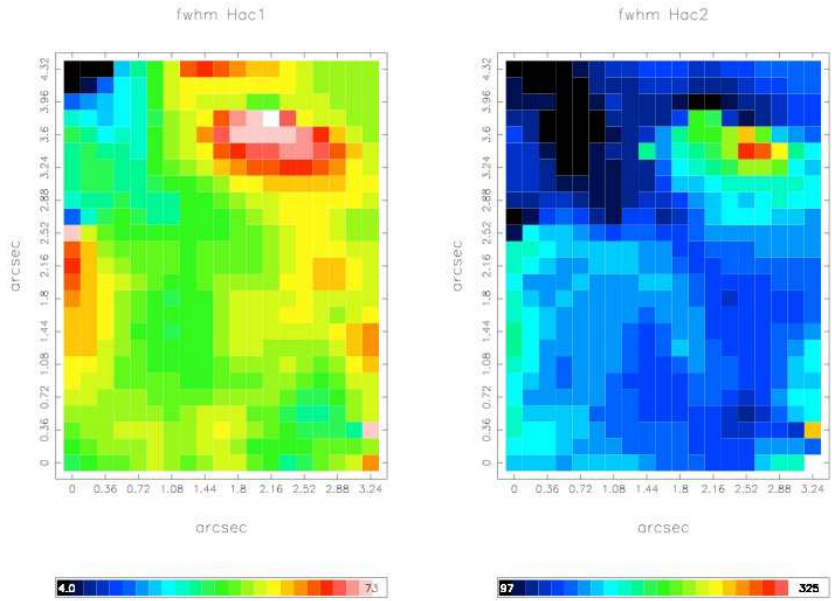}
\caption{{\bf Pos 3.} \emph{Left:} FWHM map in H$\alpha$ C1 (range 4--73); \emph{right:} FWHM map in H$\alpha$ C2 (range 97--295). A scale bar is given for each plot in \kms, with values corrected for instrumental broadening. North is up and east is left.}
\label{fig:3ha_fwhm}
\end{figure*}

The C1 intensity map (Fig.~\ref{fig:3ha_flux}, left panel) shows a bright knot in the north-west corner, with a diffuse, extended area of emission spreading out towards the south-east. This diffuse emission corresponds to part of a highly clumped, ragged structure seen on the \textit{HST} image of Fig.~\ref{fig:GMOS_finder}, which is unresolved in our IFU data. In the C2 flux map, the C1 velocity map and in both the C1 and C2 FWHM maps, this area is no longer smooth and diffuse, but rather surprisingly contains a compact region ($0\farcs8\times 1\farcs4$ $\equiv$ $9\times 16$~pc) with very different properties to the rest of the field. The FWHM of both C1 and C2 (Fig.~\ref{fig:3ha_fwhm}) peak here at 70~\kms{} and 300~\kms, respectively (corrected for instrumental broadening), both corresponding to a strong redshifted peak in C1 ($\sim$25~\kms{} with respect to $v_{\rm sys}=-80$~\kms; Fig.~\ref{fig:3ha_flux}, left panel). In Fig.~\ref{fig:pos3_overlays}, we overplot the C1 FWHM and radial velocity maps (contours) on the C1 flux map (greyscale) to illustrate how this compact region, so clearly defined in the line dynamics maps, only roughly corresponds to the north-west of the diffuse emission mentioned above, and certainly not with the bright knot in the far corner. The rapid peaks in line-width and the contrasting nature of this knot indicate that this could be the position of the SNR VLA-8. It is of approximately the right extent, but since its centre is offset by $\sim$$2''$ (based on \textit{HST} coordinates) from the location given by \citet{greve02}, we cannot confirm this scenario.

\begin{figure*}
\includegraphics[width=11cm]{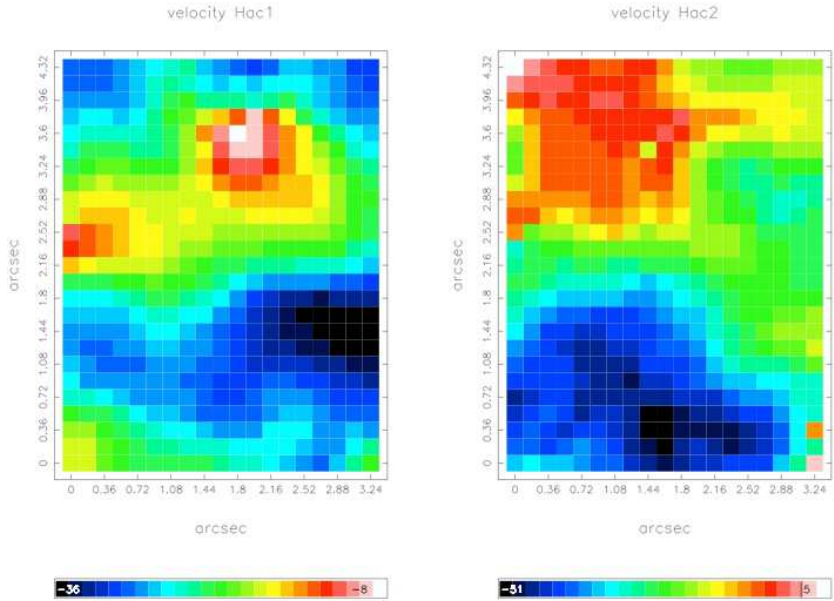}
\caption{{\bf Pos 3.} \emph{Left:} radial velocity map in H$\alpha$ C1 (range $-7$ to +21); \emph{right:} radial velocity map in H$\alpha$ C2 (range $-22$--34). A scale bar is given for each plot in \kms{} (heliocentric) corrected for the systemic velocity of the galaxy ($-80$~\kms), where zero is marked with a line.}
\label{fig:3ha_vel}
\end{figure*}
\begin{figure}
\centering
\includegraphics[width=0.475\textwidth]{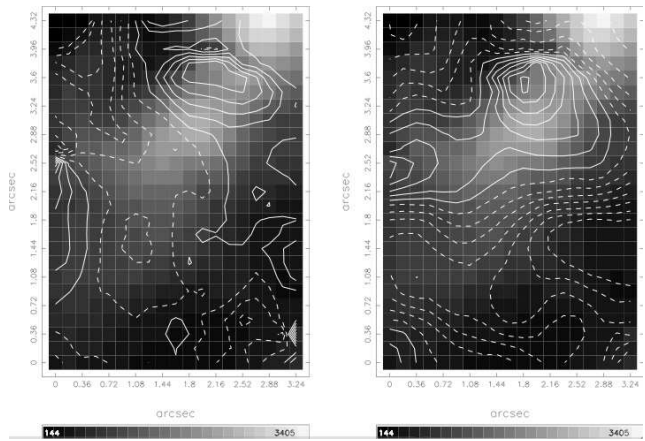}
\caption{{\bf Pos 3.} \emph{Left:} contours of C1 FWHM (12 equally spaced levels from 4 to 73~\kms) overlaid on the C1 flux map; \emph{right:} contours of C1 radial velocity (12 equally spaced levels from $-7$ to +21~\kms) overlaid on the C1 flux map. Solid contours represent levels above the mean of the range in each case, and dashed contours represent levels below the mean. These two maps illustrate the offset of the peak of broad and redshifted emission with respect to the peak in the flux distribution.}
\label{fig:pos3_overlays}
\end{figure}

A faint diagonal strip in the C2 flux map runs from the north-east to south-west of the field in the opposite direction to the C1 morphology, wrapping around the unusual knot to the north and south, and is highly correlated with the distribution of narrowest C2 line-widths. This pattern does not have a counterpart on the \textit{HST} image of Fig.~\ref{fig:GMOS_finder}. The north-eastern corner of this field is coincident with part of the evacuated region around SSC A (again see Fig.~\ref{fig:GMOS_finder}), and contains the narrowest C1 widths (as low as 5~\kms; consistent with just thermal broadening) seen in any of the four IFU fields.

\begin{figure}
\centering
\includegraphics[width=6cm]{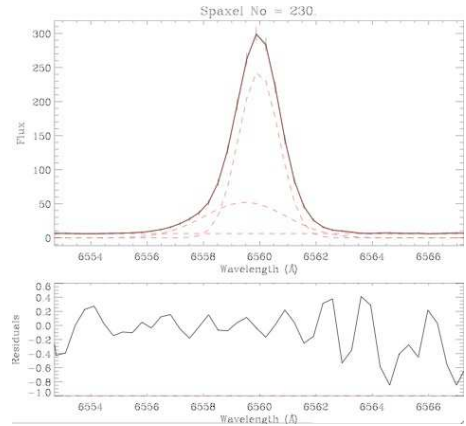}
\caption{{\bf Pos 3.} Example of our fit to the H$\alpha$ line in spaxel 230 (position marked with a circle on Fig.~\ref{fig:3ha_flux}). This spaxel exemplifies the largest velocity difference between C1 and C2 of any of the fields.}
\label{fig:3fit_egs}
\end{figure}

The gas in the south of the field, fairly unremarkable in the flux or FWHM maps, stands out in the radial velocity maps of both C1 and C2 as being very blueshifted ($\sim$10--20~\kms{} with respect to $v_{\rm sys}$; representing the most blueshifted C2 in all fields). A representative profile (together with the best-fitting model) from this region is shown in Fig.~\ref{fig:3fit_egs}. When compared to the broader context as seen in the \textit{HST} image, the blueshifted gas corresponds to an area just interior to the shell-like feature just off the south-west of the IFU FoV (described above), possibly indicating C1 and C2 are both associated with the near-side of this bubble.
 
Unlike positions 1 and 2, we do not see a correlation between C2 FWHM and C1 flux. Instead there is a strong anti-correlation between C2 FWHM and C2 flux (the maps appear almost morphologically identical), and to a lesser extent between C2 FWHM and C1 FWHM, although the respective ranges in FWHM are dramatically different. A generally faint C1 results in this field having the highest average flux ratio of C2/C1 of $\sim$30 per cent.

Electron densities were derived from the flux ratio of [S\two]$\lambda$6717/$\lambda$6731 but fall near or below the low density limit (100~\cm3) across most of the field. In order to increase S/N, we have again binned the data in a 2$\times$2 manner, and resulting electron density map for C1 is shown in Fig.~\ref{fig:3elecdens} (left panel). We find the electron density increases to a few hundred \cm3 in the north-west of the field, coincident with the strongest H$\alpha$ emission seen in Fig.~\ref{fig:3ha_flux}. The [S\two]/H$\alpha$ (map shown in Fig.~\ref{fig:3elecdens}, right panel) and [O\three]/H$\beta$ (map not shown) flux ratios are consistent with pure photoionization in every spaxel according to the `maximum starburst line' method. However, the low metallicity of NGC 1569 may mean that this threshold is an overestimate and that some contribution from shocked emission is present. Nonetheless, the low line ratios are a little unexpected considering the proximity of the SNR VLA-8.

\subsubsection{Interpretation}
\begin{figure}
\centering
\includegraphics[width=0.475\textwidth]{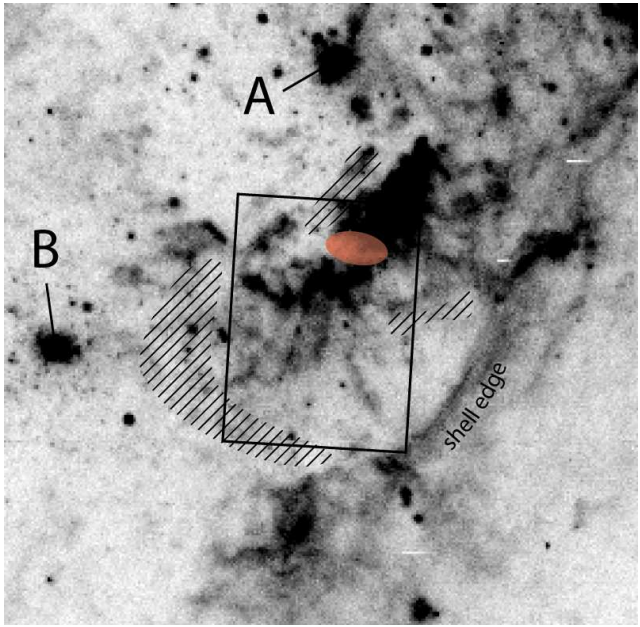}
\caption{\textit{HST} F656N image detail showing the region around IFU position 3. The black hatched areas encompass non-photoionized points found in the study of \citet{buckalew06}, and the filled ellipse marks the location of the unusual compact knot.}
\label{fig:pos3_detail}
\end{figure}

\begin{figure*}
\begin{minipage}{5.5cm}
\includegraphics[width=5.5cm, height=7.5cm]{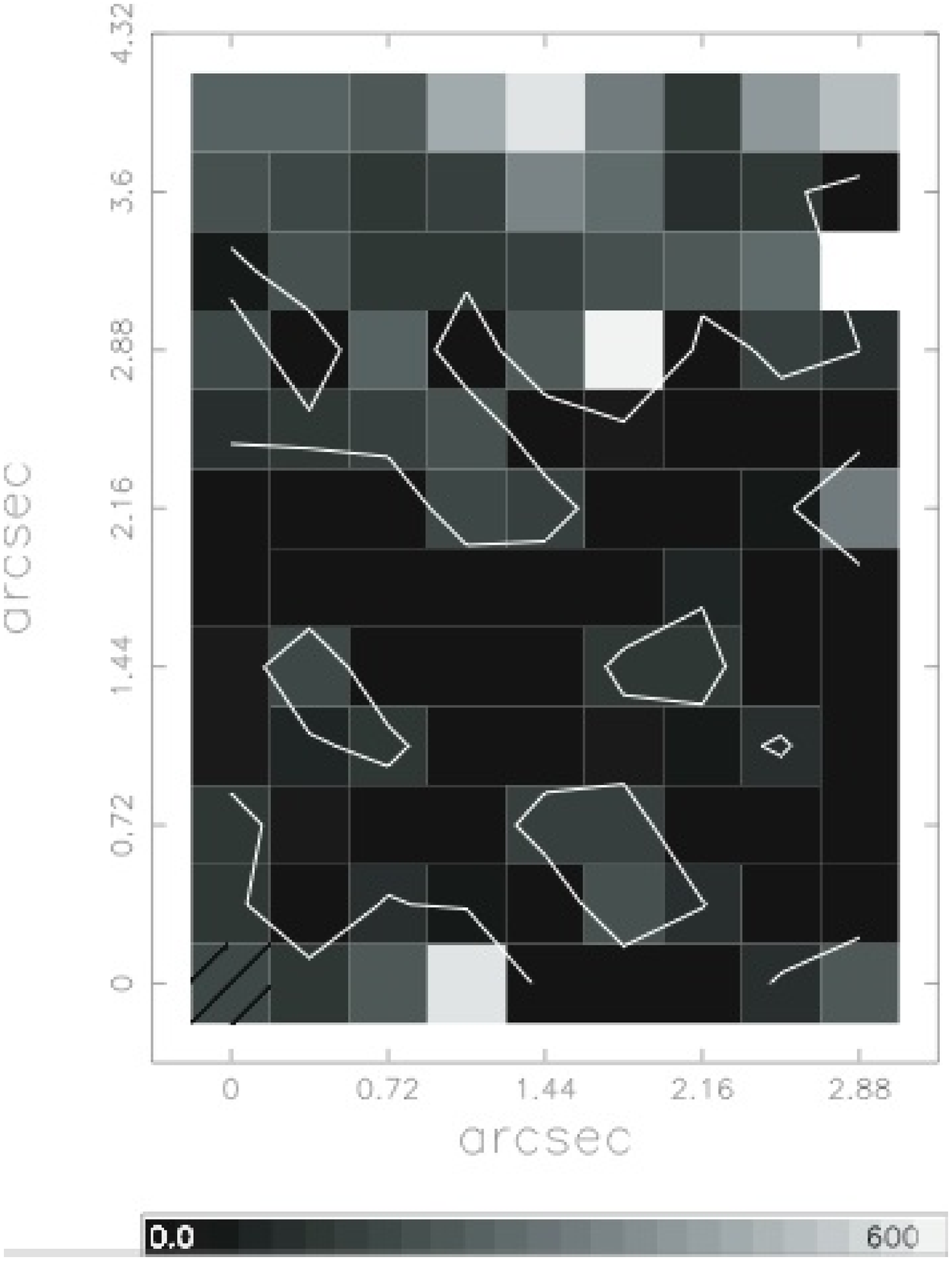}
\end{minipage}
\hspace*{0.8cm}
\begin{minipage}{5.5cm}
\includegraphics[width=5.5cm, height=7.5cm]{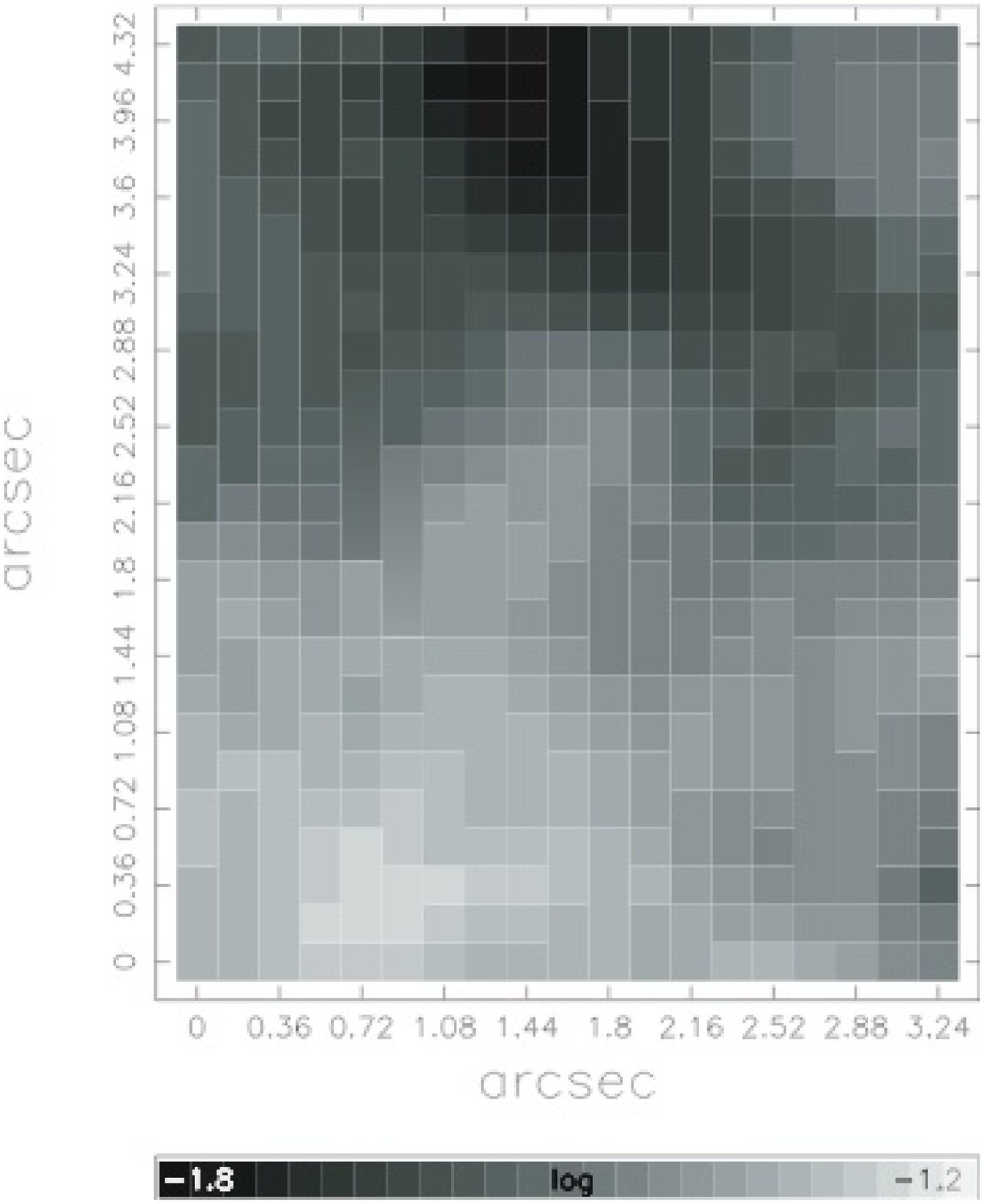}
\end{minipage}
\caption{{\bf Pos 3.} \emph{Left:} Electron density map derived from the ratio of [S\two]$\lambda6717 / \lambda6731$ for C1 [a scale bar is given in units of cm$^{-3}$; a single contour level marks the 100~cm$^{-3}$ low density limit \citep{osterbrock89}]; \emph{right:} ratio of [S\two]$\lambda$6717+$\lambda$6731/H$\alpha$ in logarithmic scale.}
\label{fig:3elecdens}
\end{figure*}

The striking feature in position 3 is the unusual compact knot that dominates the C1 width and velocity maps, and the C2 flux and width maps. Fig.~\ref{fig:pos3_detail} shows a magnified version of the region surrounding this IFU position on which the location of the compact knot is shown by a filled ellipse. As we have suggested, the sudden peak in line-widths could indicate that this is the location of the SNR VLA-8 catalogued by \citet{greve02}, but since we see no evidence for shocked line-ratios (as would be expected if this was a SNR), and the fact that it is offset from the position quoted for VLA-8 make this explanation less tenable. Ergo, we can only conclude that this feature represents an isolated, unusually turbulent knot of faintly emitting, redshifted gas that happens to lie on this sight-line.

Except for at the location of this knot, the width of C2 does not exceed $\sim$200~\kms{} over the whole field, and falls as low as 100~\kms{} in the north-east quadrant. Away from the knot, the broadest widths are not associated with brightly emitting gas (as found in positions 1 and 2) implying that we are not seeing a strong interaction with the cluster winds in this position. The average ratio of C2/C1 for this position is the highest of all the fields, giving another indication that the physical conditions in this field are different to that of the other positions. 

It would seem contradictory why at this close projected distance from SSCs A and B ($\sim$50~pc) the gas has not completely been blown away, since large swathes of material at similar projected distances have been completely evacuated. Can we explain this using the observations? We have already mentioned that we see little evidence for a clump--wind interaction in the form of broad line-widths or shocked line-ratios. Both C1 and C2 become gradually blueshifted with respect to $v_{\rm sys}$ towards the south of this field, peaking in an area coincident with the interior of the shell-like feature located just off the bottom of the field, if the string of non-photoionized points identified by \citet{buckalew06} are included as part of the shell edge. Their non-photoionized points are shown in Fig.~\ref{fig:pos3_detail} as hatched areas, and indicate that the eastern edge of this shell, although not H$\alpha$ bright, is shock-excited. This whole region, with the exception of the unusual isolated knot, may therefore be located at a different line-of-sight to the surrounding gas and be shielded to a certain extent from the effects of the strong cluster winds. This would explain the relatively quiescent line-widths, and was a result predicted for the turbulent mixing/ablation model described in \citetalias{westm07a}, for regions not subject to the direct impact of cluster winds.

\subsection{GMOS Position 4} \label{sect:GMOS_pos4}

Position 4 covers a bright knot of ionized gas to the east of SSC A that has a morphology somewhat resembling the southern knot in position 1. This condensation is coincident with the radio continuum source M-5 \citep[a ``non-thermal source with some thermal contribution'';][]{greve02}, and appears to have an oval shell-like extension to the south in the \textit{HST} F656N image (Fig~\ref{fig:GMOS_finder}). 

\begin{figure*}
\includegraphics[width=16cm]{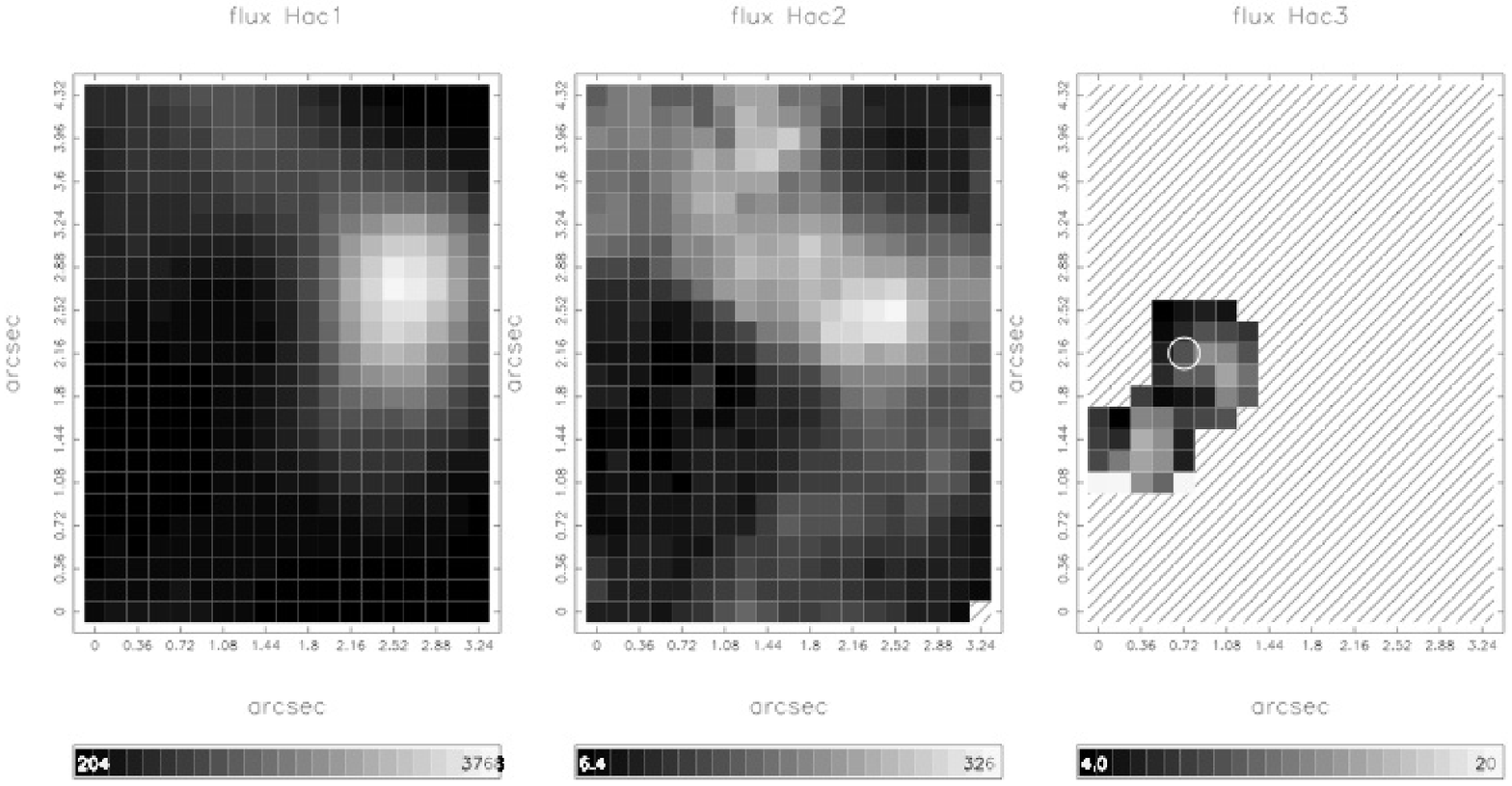}
\caption{{\bf Pos 4.} \emph{Left:} Flux map in H$\alpha$ C1 (range 204--3768); \emph{centre:} flux map in H$\alpha$ C2 (range 6.4--326); \emph{right:} flux map in H$\alpha$ C3 (range 4.0--20). Non-detections are represented as hatched spaxels, the $x$ and $y$ scales are in arcseconds offset from the lower-left spaxel, and a scale bar is given for each plot with units $10^{-15}$ erg s$^{-1}$ cm$^{-2}$~spaxel$^{-1}$. North is up and east is left. The spaxel from which the example H$\alpha$ profile shown in Fig.~\ref{fig:4fit_egs} was extracted is marked with a circle.}
\label{fig:4ha_flux}
\end{figure*}

\begin{figure*}
\includegraphics[width=16cm]{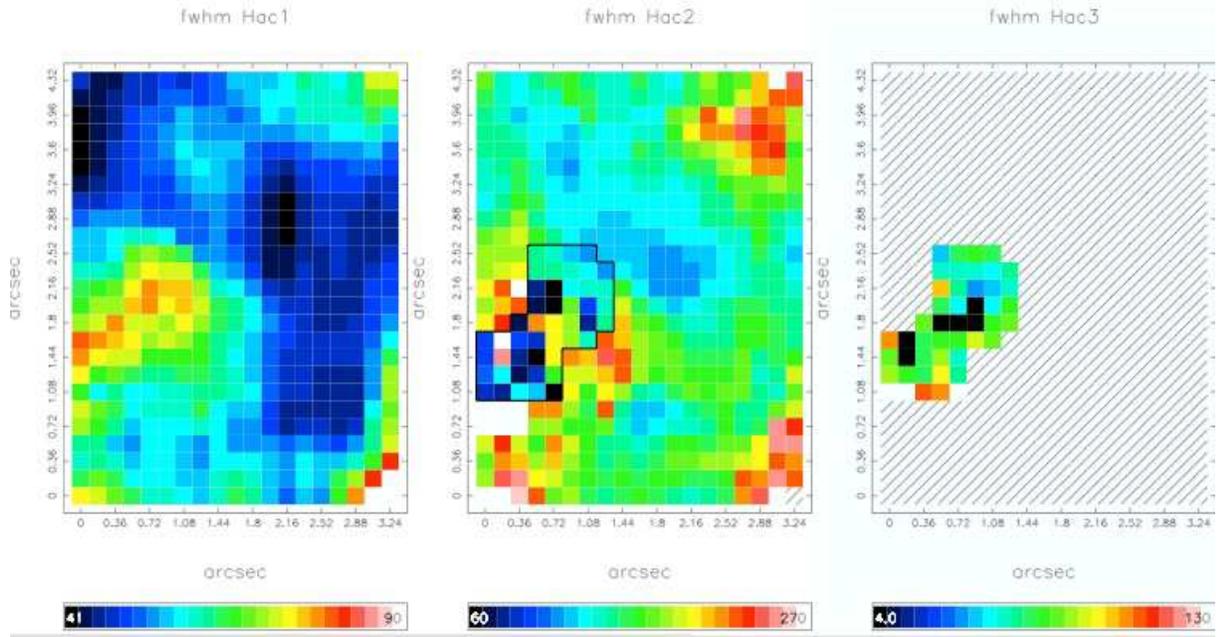}
\caption{{\bf Pos 4.} \emph{Left:} FWHM map in H$\alpha$ C1 (range 41--90); \emph{centre:} FWHM map in H$\alpha$ C2 (range 60--270); \emph{right:} FWHM map in H$\alpha$ C3 (range 4--130). The region outlined with a solid line in the C2 map indicates where the measurement errors are high due to the addition of a third Gaussian to the fit. A scale bar is given for each plot in units of \kms, with values corrected for instrumental broadening. North is up and east is left.}
\label{fig:4ha_fwhm}
\end{figure*}

\begin{figure*}
\includegraphics[width=16cm]{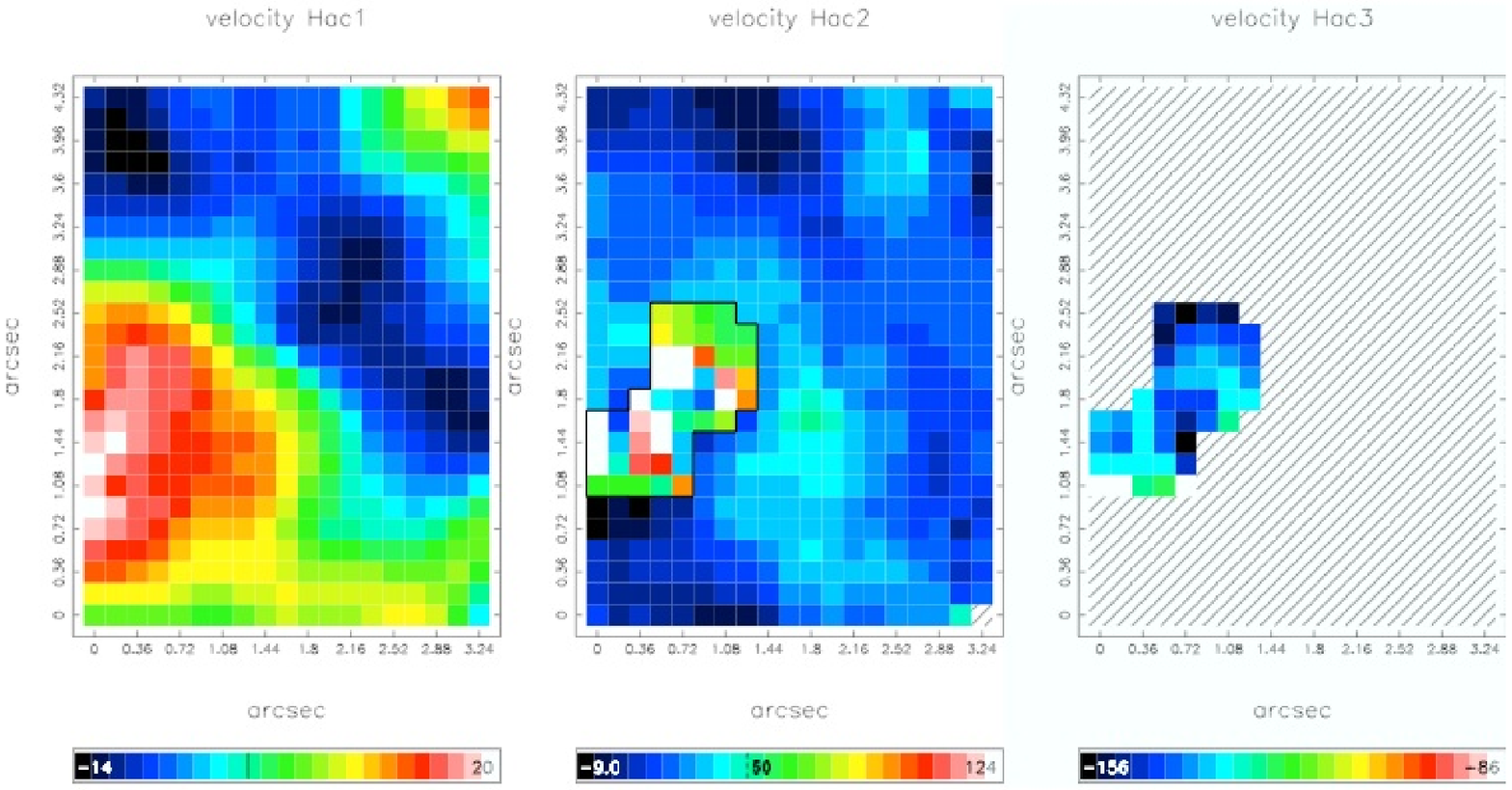}
\caption{{\bf Pos 4.} \emph{Left:} radial velocity map in H$\alpha$ C1 (range $-14$--20); \emph{centre:} radial velocity map in H$\alpha$ C2 (range $-9$--124); \emph{right:} radial velocity map in H$\alpha$ C3 (range $-156$ to $-86$). Again, the region outlined with a solid line in the C2 map indicates where the measurement errors are high due to the addition of a third Gaussian to the fit. A scale bar is given for each plot in \kms{} (heliocentric) corrected for the systemic velocity of the galaxy ($-80$~\kms), where zero is marked with a line in the left-hand panel and +50~\kms\ is marked with a dashed line in the central panel.}
\label{fig:4ha_vel}
\end{figure*}

\begin{figure}
\centering
\includegraphics[width=6cm]{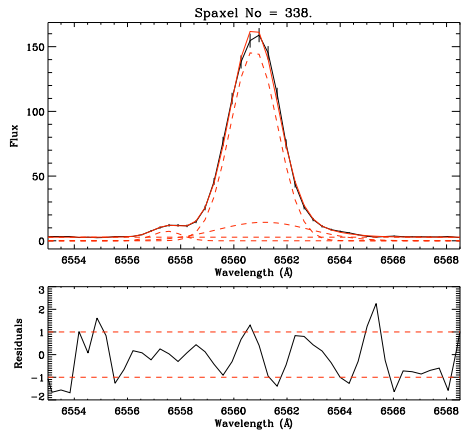}
\caption{{\bf Pos 4.} Example of our fit to the H$\alpha$ line in spaxel 338 (position marked with a circle on Fig.~\ref{fig:4ha_flux}). This example was chosen to illustrate the presence of three significant line components found in the area on the east of the field. The third component, C3, is the small profile in the blue wing.}
\label{fig:4fit_egs}
\end{figure}

Before discussing the emission-line maps for this position (Figs~\ref{fig:4ha_flux}--\ref{fig:4ha_vel}), we first comment that since a third component is required in a small region on the east of the field, its presence has resulted in high uncertainties on the fits to C2 in this same region \citepalias[for reasons discussed in][]{westm07a}. This has caused a spurious and erratic region in the C2 FWHM and radial velocity maps (outlined with a solid line on the following maps for clarity), so we ignore this section when discussing the properties of C2. Fig.~\ref{fig:4fit_egs} shows an example H$\alpha$ line profile from this region showing the best-fitting Gaussian model profiles and the fit residuals. In this particular spaxel, a third component is clearly required to accurately model the integrated line shape, but it is clear how the associated errors on the fit to C2 would increase as C3 becomes fainter thus leading to this area of inconsistent results.


The aforementioned bright knot shows up clearly in the C1 flux map (Fig.~\ref{fig:4ha_flux}, left panel) as expected, but is only evident in the corresponding FWHM and radial velocity maps and the C2 flux map as part of a distinct arc of gas extending out of the FoV the north-east, and partly towards the south-east. This arc has a consistently low C1 FWHM ($\sim$40~\kms; Fig.~\ref{fig:4ha_fwhm}, left panel), and is blueshifted by $-10$ to $-15$~\kms{} with respect to $v_{\rm sys}$ (Fig.~\ref{fig:4ha_vel}, left panel), implying that the bright knot is dynamically associated with a larger gas filament extending out either side of the FoV.

The gas not associated with the filament is much fainter, and exhibits very different dynamical characteristics. A large circular region in the east of the field is distinct in C1 for having coincident peaks of very broad (up to $\sim$90~\kms; representing the broadest C1 of all fields) and redshifted velocities (up to +20~\kms{} relative to $v_{\rm sys}$). Around this location a third component is also required to fit the line shape (as mentioned above). From the velocity difference between C1 and C3, we infer expansion velocities of order $v_{\rm exp}\approx 80$~\kms, with the fainter component of the two (C3) travelling at $\sim$150~\kms{} towards us. An absence of increased line-widths or a third component at the location of the small shell seen to the south of the knot in the \textit{HST} H$\alpha$ image indicates that this structure may not be expanding, is expanding only very slowly, or not a shell at all. The broadest C2 lines ($\sim$250~\kms; Fig.~\ref{fig:4ha_fwhm}, centre panel) are found in the south-east and north-west, and in both locations are associated with redshifted gas in C1 (+5--20~\kms). We mention this because these are sections of the IFU FoV that appear to be coincident with part of the evacuated cavity surrounding SSC A, and will be discussed in Section~\ref{sect:cavity}.

This field has a number of unexpected relationships between the components compared to the other positions. Firstly, this is the only position where the flux distribution of C2 (Fig.~\ref{fig:4ha_flux}, centre panel) bears any resemblance to the corresponding C1 map and to the \textit{HST} image. We show below how this can be explained by the unusually high density of the gas emitting C2. Secondly, the correlation between C2 FWHM and C1 flux is in the opposite sense to what we have seen in positions 1 and 2. Here the broadest C2 lines are coincident with the \emph{faintest} C1 emission in the north-west, south-west and centre-east of the field.

\begin{figure*}
\begin{minipage}{11cm}
\includegraphics[width=11cm,height=8cm]{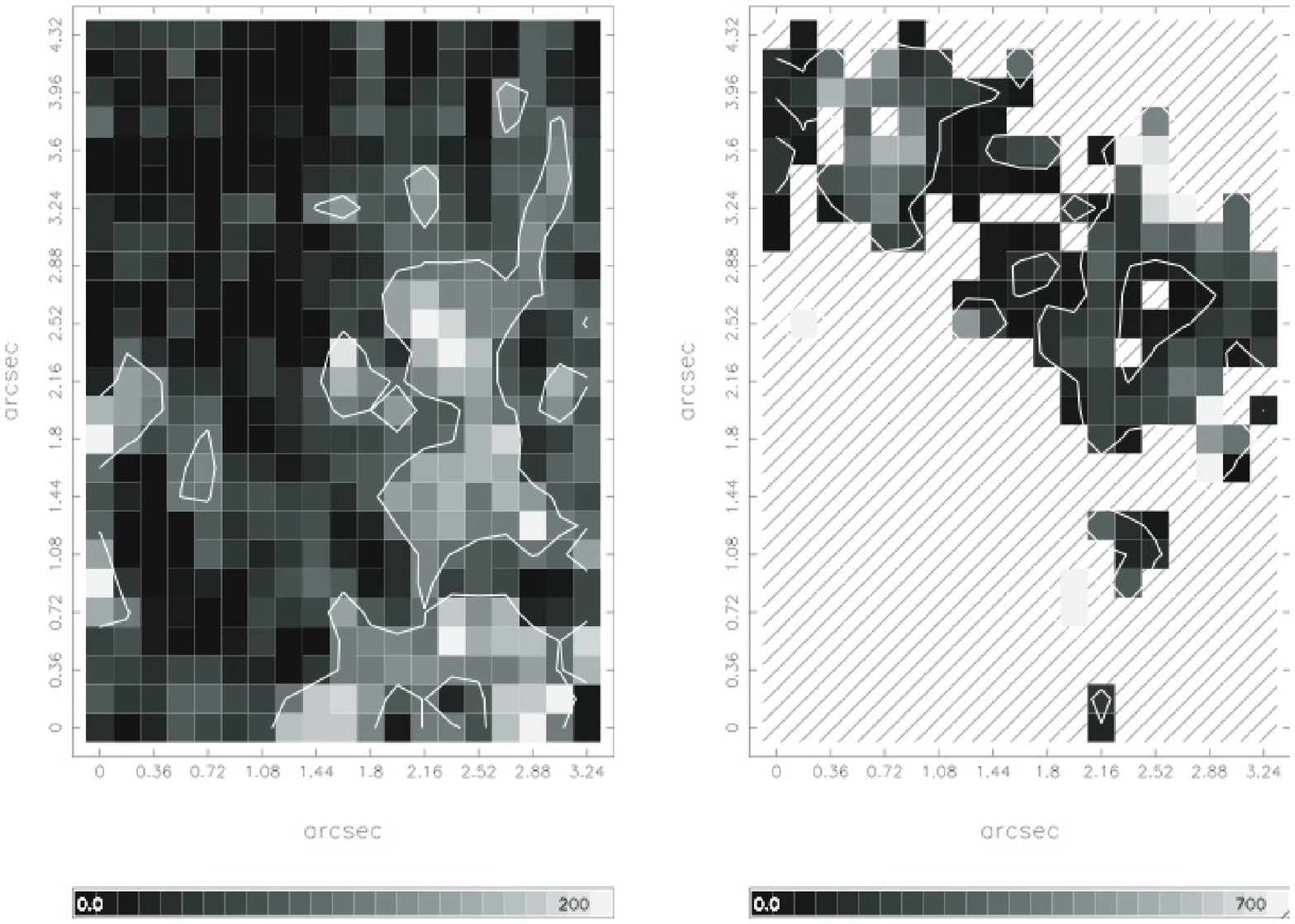}
\end{minipage}
\hspace*{0.3cm}
\begin{minipage}{5.8cm}
\includegraphics[width=5.5cm,height=8cm]{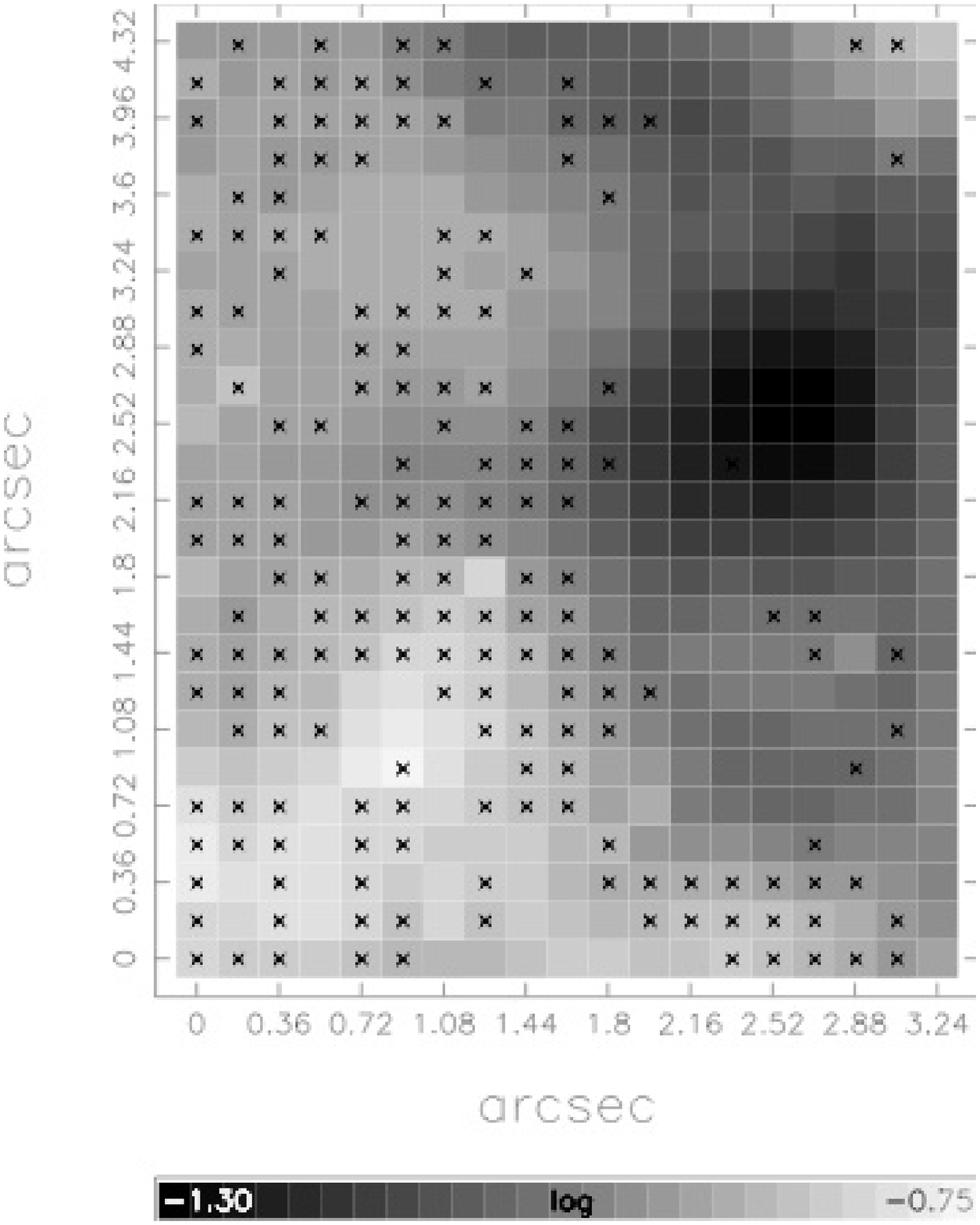}
\end{minipage}
\caption{{\bf Pos 4.} \emph{Left:} Electron density map derived from the ratio of [S\two]$\lambda$6717/$\lambda$6731 for C1 (range 0--200); \emph{centre:} electron density map for C2 (range 0--700). A scale bar is given for each plot in units of cm$^{-3}$; a single contour level marks the 100~cm$^{-3}$ low density limit \citep{osterbrock89} on both maps. \emph{Right:} flux ratio of log([S\two]($\lambda$6717+$\lambda$6731)/H$\alpha$) C1 (log range $-1.3$ to $-0.75$). Spaxels marked with a cross indicate non-photoionized emission according to the \citet{kewley01} maximum starburst line.}
\label{fig:4elecdens}
\end{figure*}

This is the only position for which we have been able to make confident double-Gaussian fits to some of the [S\two] lines, and derive the electron density for both components (mapped in Fig.~\ref{fig:4elecdens}). The density of the gas emitting C1 (left panel) is low, with only parts of the western edge coincident with the bright knot rising above the low density limit (indicated by the single contour). We are able to derive the density for C2 (centre panel) in a diagonal band stretching from the centre-west to the north-east, coincident with the band of bright C2 emission and blueshifted C1 gas in the north of the field. The density of C2 is on average much higher than that of C1, reaching up to $\sim$600~\cm3 in the north-east. Since the emission intensity of recombination lines is proportional to the gas density-squared, the unusually high densities in C2 may may explain why in this field the C2 emission morphology bears a close resemblance to the corresponding C1 morphology.

Also shown in Fig.~\ref{fig:4elecdens} is the flux ratio of [S\two]($\lambda$6717+$\lambda$6731)/H$\alpha$ (right panel). The lowest ratios are found in the centre-west of the field, coincident with the bright H$\alpha$ knot and the radio source M-5. Using the [O\three]/H$\beta$--[S\two]/H$\alpha$ nebular diagnostic technique described in \citetalias{westm07a} and above, we find a significant number of spaxels over the whole eastern half of the field contain a possible contribution from non-photoionization processes \citep[presumably shock excitation; see][]{calzetti04}. However, the derived low flux ratios at the location of the bright knot do not support a SNR origin for M-5 unless it is highly embedded.

\subsubsection{Interpretation}
Evident in both the flux maps of C1 and C2, a band of material curves round from the north-east of the field to the south, including the bright knot near the western edge. In C1, this band has a consistently low FWHM and is blueshifted by 10--15~\kms{} with respect to $v_{\rm sys}$. In comparison, the ionized gas to the east of the field is significantly redshifted in C1, has a high-velocity ($\sim$150~\kms) blueshifted component, and a widespread non-photoionized contribution to the excitation. 

\begin{figure}
\centering
\includegraphics[width=0.475\textwidth]{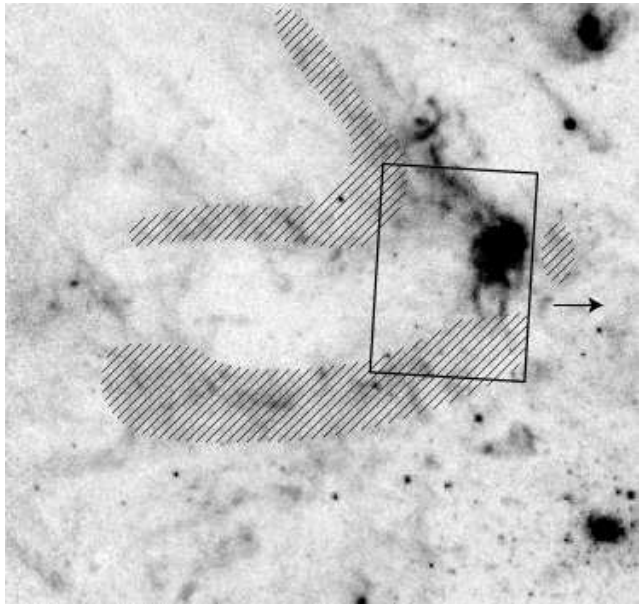}
\caption{\textit{HST} F656N image detail showing the region around IFU position 4. The hatched areas are non-photoionized points found in the study of \citet{buckalew06}, and the arrow indicates the direction of SSC A.}
\label{fig:pos4_detail}
\end{figure}

Careful inspection of the \textit{HST} image of Fig.~\ref{fig:GMOS_finder} shows that this band may be the westernmost part of a fragmented shell extending out of the FoV to the east. Fig.~\ref{fig:pos4_detail} shows a blow-up of the \textit{HST} F656N image around position 4. The hatched areas represent the regions of non-photoionized points found by \citet{buckalew06}, illustrating how the edges of this shell appear to be traced out by non-photoionized (shocked) emission. How can we relate this to our more wide-spread distribution of non-photoionized gas shown in Fig.~\ref{fig:4elecdens}? The methods used by \citeauthor{buckalew06} for the determination of their points are inherently luminosity weighted, in the sense that only where they actually detected emission at sufficient S/N could they measure the line ratios. The fact that our data are significantly deeper, combined with the advantages of actually being able to fit the individual emission lines (thus avoiding the uncertainties of continuum subtraction), has meant that we have been able to accurately measure the line ratios for much fainter gas. The distribution of non-photoionized gas we derive includes their points arising from the brighter regions and additional points identified from the lower surface-brightess regions, implying that shock-excited emission is more widespread than previously observed.

We can now associate the emission to the east of the field with originating from within the expanding shell structure, where the near-side wall is represented by the high-velocity C3, and the far-side by the brighter redshifted region seen in C1. The velocities and brightness contrast between the two components can be explained if the far-side of the bubble is expanding into a dense medium, causing it to slow down rapidly and emit brightly. The fast velocities and faint H$\alpha$ emission of C3, imply that the near-side is expanding relatively unimpeded, meaning that an estimate of the true expansion velocity of the bubble can be made by comparing the radial velocity of the gas in the shell edge (the blueshifted band) to the high-velocity C3. We estimate $v_{\exp}$ to be $\approx$\,140~\kms{} by taking half the velocity difference between the two line components.

The width of C2 remains between 100 and 150~\kms{} in the blueshifted shell edge including the position of the bright knot. This lack of very broad lines may indicate that a direct wind--clump interaction is not taking place here. Recall that this clump contains the radio source M-5 emitting a mixture of thermal and non-thermal radiation. The presence of a thermal signature points towards an embedded photoionizing source (e.g.~a young star cluster) being the origin of this emission. Emission in the thermal IR would support this scenario, but unfortunately the high-resolution observations of \citet{tokura06} did not cover this knot. There is very marginal evidence for continuum emission at the location of the bright knot in our IFU data, but since we cannot identify any discrete stellar features (e.g.~a WR bump) we cannot prove the existence of an embedded stellar source.

\section{The state of the ISM} \label{sect:ISM_state}
\begin{figure*}
\includegraphics[width=12cm]{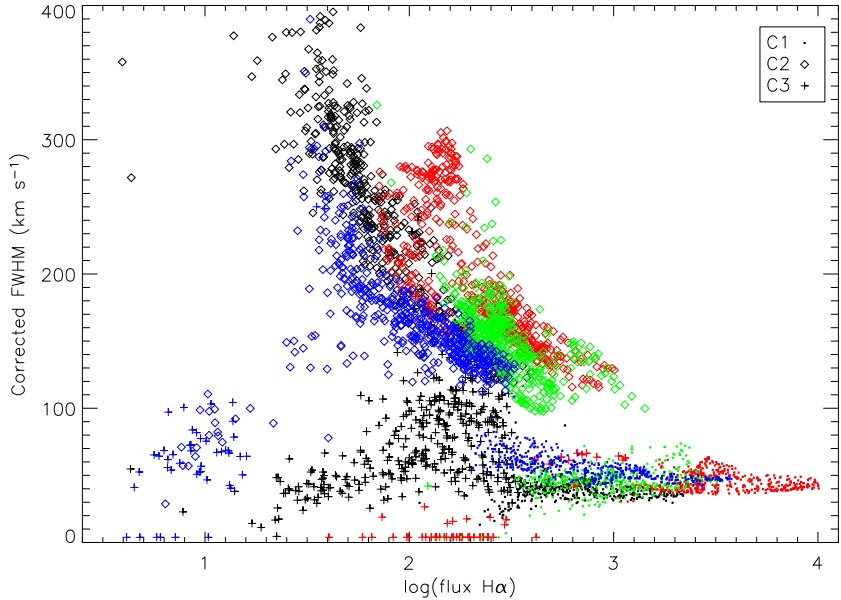}
\caption{FWHM (corrected for instrumental broadening) vs.~log(flux H$\alpha$) (arbitrary scale) for all four IFU positions.  Black symbols represent position 1, red position 2, green position 3, and blue position 4.}
\label{fig:sigma_flux}
\end{figure*}

\begin{figure*}
\includegraphics[width=12cm]{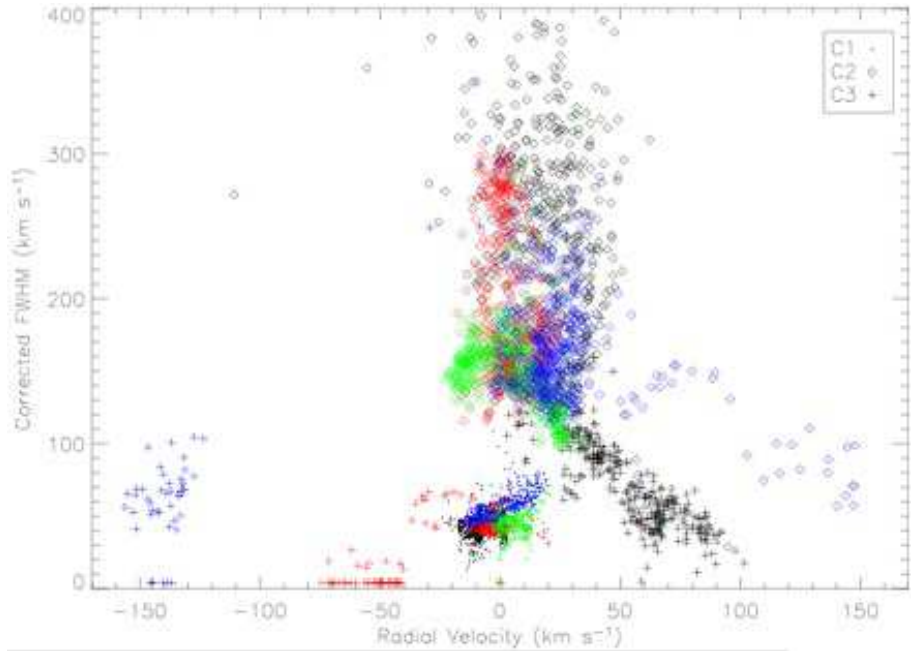}
\caption{FWHM (corrected for instrumental broadening) vs.~heliocentric radial velocity (with respect to $v_{\rm sys}$) for all four IFU positions. Symbol colours are the same as Fig.~\ref{fig:sigma_flux}.}
\label{fig:sigma_vel}
\end{figure*}

\begin{figure*}
\includegraphics[width=12cm]{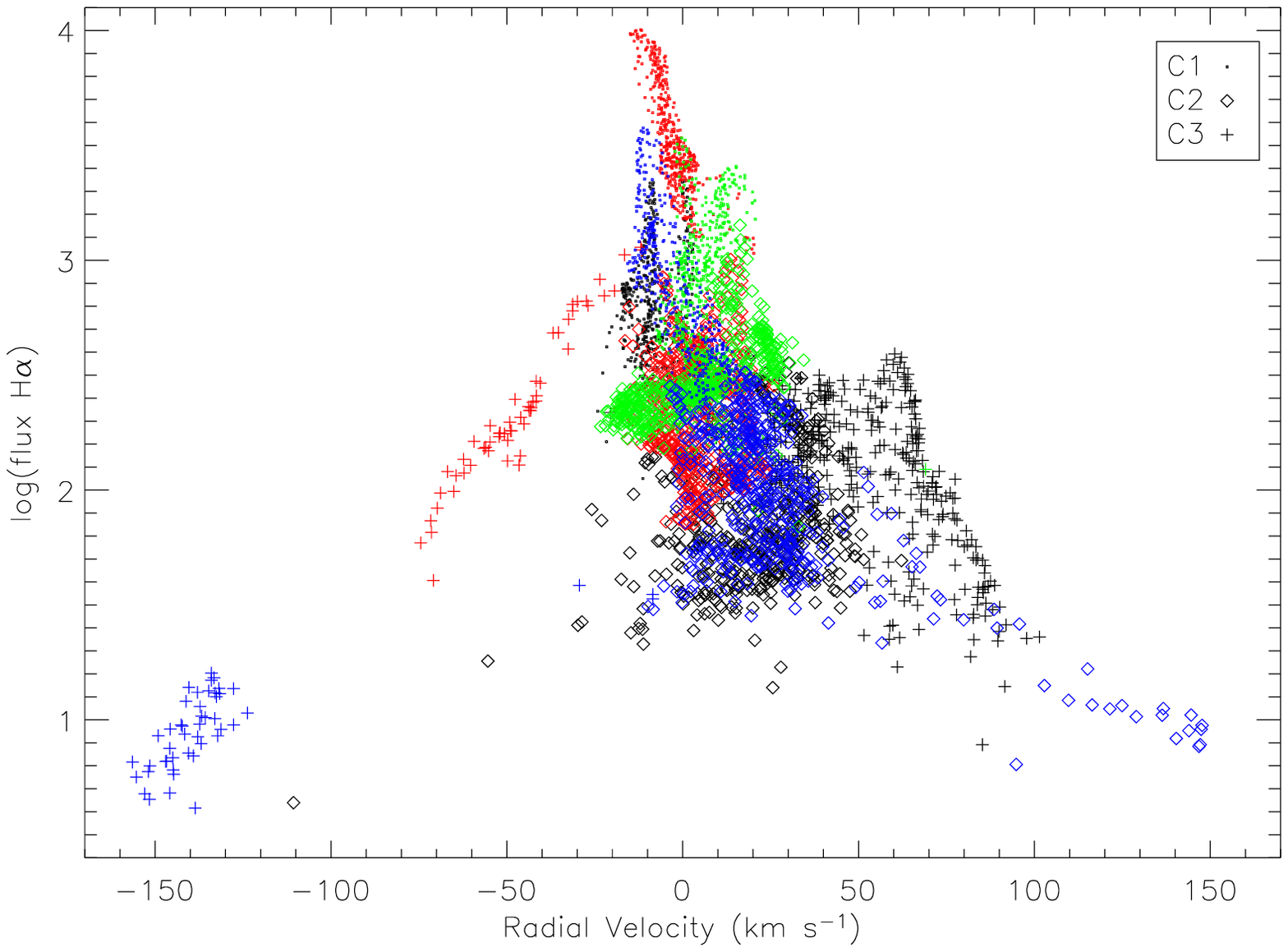}
\caption{Heliocentric radial velocity (with respect to $v_{\rm sys}$) vs.~log(flux H$\alpha$) (arbitrary scale) for all four IFU positions. Symbol colours are the same as Fig.~\ref{fig:sigma_flux}.}
\label{fig:vel_flux}
\end{figure*}

So far we have examined the details of the ionized gas distribution in each field individually. However, there are insights to be gained from looking at the collective ionized gas properties of all four fields at once, so as to assess the general trends and systematic variations across the $\sim$200~pc wide region centred on SSC A as sampled by our IFU pointings. In Figs~\ref{fig:sigma_vel}, \ref{fig:sigma_flux} and \ref{fig:vel_flux}, we plot the Gaussian flux, FWHM and radial velocity of the identified line components for all four positions in the three possible permutations. In the following paragraphs, we will discuss these plots together with the selected average properties listed in Table~\ref{tbl:averages}, in the context of understanding the general state of the ISM.

\begin{table}
\begin{center}
\caption{Selected average (mean) emission-line properties for the four IFU fields. The velocity measurements are quoted with a error of $1\sigma$ on the mean, and do not take into account uncertainties on the individual data points.}
\label{tbl:averages}
\begin{tabular}{l l l r@{\,$\pm$\,}l}
\hline
IFU & Line & Gaussian property & \multicolumn{2}{c}{Average value} \\
position & component \\
\hline
all & C2/C1 & Flux ratio & \multicolumn{2}{c}{0.12} \\
1 & C2/C1 & Flux ratio & \multicolumn{2}{c}{0.11} \\
2 & C2/C1 & Flux ratio & \multicolumn{2}{c}{0.07} \\
3 & C2/C1 & Flux ratio & \multicolumn{2}{c}{0.30} \\
4 & C2/C1 & Flux ratio & \multicolumn{2}{c}{0.13} \\
all & C1 & FWHM\,$^{a}$ & 46.9 & 8.7 \\
all & C2 & FWHM & 192.0 & 47.3 \\
all & C3 & FWHM & 53.1 & 30.0 \\
all & C1 & Radial Velocity\,$^{b}$ & $-1.8$ & 6.7 \\
all & C2 & Radial Velocity & $+13.4$ & 17.4 \\
all & C3 & Radial Velocity & $-14.6$ & 24.3\,$^{c}$ \\
1 & C1 & Radial Velocity & $-16.7$ & 6.0 \\
2 & C1 & Radial Velocity & $-2.5$ & 5.5 \\
3 & C1 & Radial Velocity & $-3.3$ & 5.9 \\
4 & C1 & Radial Velocity & $-0.1$ & 9.3 \\
1 & C2 & Radial Velocity & $+19.5$ & 21.9 \\
2 & C2 & Radial Velocity & $+4.4$ & 8.7 \\
3 & C2 & Radial Velocity & $+4.4$ & 14.4 \\
4 & C2 & Radial Velocity & $+25.3$ & 24.7 \\
\hline
\end{tabular}
\begin{tabular}{p{8cm}}

$^{a}$ FWHMs are in units of \kms{}, corrected for instrumental, but not thermal broadening. \\
$^{b}$ Radial velocities are in units of \kms{}, relative to $v_{\rm sys}$ ($= -80$~\kms). \\
$^{c}$ Average without position 4 = $+24.5\pm 20.6$~\kms
\end{tabular}
\end{center}
\end{table}


Plots similar to Fig.~\ref{fig:sigma_flux} have been used successfully in the past as diagnostic tools to examine line broadening in giant H\two\ regions \citep{m-t96, martinez07}. These studies find inclined bands of points within the FWHM--flux plots to be associated with the presence of large-scale expanding shells within the H\two\ region, and use their presence to constrain the evolutionary stage of the region. We note that similar inclined bands are observed in our plot, but caution that unlike the aforementioned studies, we decompose the integrated line profiles rather than fitting only a single Gaussian. The variation in width of the single fitted Gaussian is one of the main ideas on which their conclusions are based, making a direct comparison with their conclusions difficult.

Fig.~\ref{fig:sigma_flux} shows that the trend found in position 2, whereby the broadest emission is the faintest, also applies to the entire data-set. This implies that the most turbulent gas does not emit strongly in H$\alpha$. Plotting all the data together also highlights the fact that the highest velocity gas is also the faintest (Fig.~\ref{fig:vel_flux}). On average, position 2 contains the brightest H$\alpha$ emission (as would be expected from the \textit{HST} H$\alpha$ image, Fig.~\ref{fig:GMOS_finder}), whereas position 1 is the faintest. However, position 1 exhibits the broadest C2 widths, with an even spread extending up to $\sim$400~\kms. As we have previously mentioned, this is likely to indicate that this position includes gas that is experiencing a strong interaction with the cluster winds. Position 3 has the most quiescent C2 widths, most likely resulting from the fact that the gas in this field is shielded to some extent from the direct influence of the outflowing cluster winds.

For position 2 \citepalias{westm07a}, the lower limit found for the width of C1 (referred to as FWHM$_{0}$), and its similarity to the integrated H\one{} velocity dispersion for the whole of NGC 1569  \citep[FWHM $\approx$ 35~\kms;][]{muhle05}, led us to conclude that it represents a convolution between the disturbed, turbulent ISM that has been stirred up by the effects of the intense star-formation and supernova shocks and gravitational virial motions. We could not rule out an additional minor contribution to its width from inevitable unresolved kinematical components along the line-of-sight. By plotting the FWHM results for all fields together (Figs~\ref{fig:sigma_vel} and \ref{fig:sigma_flux}), we find this consistency in C1 line-width is mirrored across all positions, and that they fall in a very narrow band ($47\pm 9$~\kms; Table~\ref{tbl:averages}) extending over the full range of intensities observed. The width of C1 is therefore not be related to its emission strength. The clear lower-limit to the FWHM of C1 found in position 2 is also evident in the other fields. In positions 1 and 3, FWHM$_{0} \approx 30$--35~\kms{} (although is less well defined for position 3), and in position 4, FWHM$_{0}$ is slightly higher at $\approx$\,45~\kms. This level of consistency therefore supports our conclusions about C1 reached in \citetalias{westm07a}. Although in places, C1 is associated with fast motion and/or expansion, its line-width is everywhere equal to or greater lower limit, FWHM$_{0}$, of the gas across the whole central regions.

We can identify a third component in positions 1, 2 and 4, and from Figs~\ref{fig:sigma_vel} and \ref{fig:sigma_flux}, see that its FWHM is equivalent to that of C1 (within the uncertainties) in every case. This confirms the conclusion that where identified, C3 represents an additional narrow component associated with a secondary region of ionized gas, possibly one half of a wind-blown bubble. In position 4 its relation to expanding shells is clear, but in position 1 its significance is not so obvious. This component occupies a prominent diagonal band in Fig.~\ref{fig:sigma_vel}, distinct from all the other components, between velocity = +100~\kms, FWHM = 20--30~\kms, and velocity = 30~\kms, FWHM = 100~\kms, implying that bulk motion and turbulence influence the dynamics of this component to differing degrees from one half of the field to the other. This behaviour is mirrored in Figs~\ref{fig:sigma_flux} and Fig.~\ref{fig:vel_flux}, where C3 of position 1 again occupies its own unique space. Its FWHM seems correlated with that of C2 at the position of the southern knot, implying that it is affected by the wind--clump interaction (unlike C1), however its velocity is spatially correlated with C1 only at the position of the northern knot. Consequently, we conclude that in position 1, C3 represents a mix of overlapping kinematical systems originating from separate mechanisms. Where it exhibits a high comparative redshift (in the north), it may represent the far-side of an expanding structure -- its properties are consistent with being part of the high-velocity kinematic subsystem identified by \citet{tomita94} and \citet{heckman95}. Where its velocity is similar to C1 (in the south), it is likely to be associated with the wind--clump interaction, since its FWHM is correlated with that of C2 (see \citetalias{westm07a}). In summary, our C3 data reveal hidden shells that are not found from morphology alone. Most of these have modest expansion velocities (30--40~\kms), which are enough to stir up the ISM but not drive a wind or produce a hot ISM.

As listed in Table~\ref{tbl:averages}, the average radial velocity of C2 is offset from the corresponding C1 average by only $\sim$+15~\kms{} over the four fields; the largest individual difference being in position 1 of $\sim$35~\kms{} for reasons discussed above. Furthermore, ignoring the high C3 velocities in positions 1 and 4, the total velocity spread between the components across the four fields is only $\sim$70~\kms{} (see e.g.~Fig.~\ref{fig:sigma_vel}). These values are on the order of the narrowest line-widths measured, and are all much less than sound speed in the hot medium ($\sim$500~\kms\ at $\sim$$10^{7}$~K), the escape velocity \citep[$\sim$75--100~\kms;][]{martin98}, and the inferred velocity of the wind. This implies that turbulent velocities dominate over bulk motions in this whole 200~pc central region.

We began this study with the aim of looking for evidence of the roots of the galactic wind, and to try and identify where and how the flow is accelerated and collimated. \citet{heckman95} and \citet{martin98} identified and catalogued large-scale expanding supershells surrounding the central regions of NGC 1569 that together form the large-scale galactic outflow. These authors find the velocity ellipses associated with shells A, B (on the western side) and E (on the north-eastern side) to have inner boundaries located at $\sim$$10''$, $\sim$$30''$ and $\sim$$60''$ away from SSC A, respectively, placing the start of the collimated flows at least $\sim$100~pc away from our IFU positions. We know that within the central 200~pc (as sampled by our observations), we do not see evidence for bulk motions consistent with the outflow. This must therefore mean that the collimation process occurs between radii of 100 to 200~pc from SSC A. This interesting region clearly deserves further detailed study, as it must be very important in the galactic outflow formation process.

Are we therefore detecting the galactic wind roots in our IFU positions? In \citetalias{westm07a} we discussed how our observations are reminiscent of the simulated gas motions within the outer bounding shock of an expanding zone as modelled by \citet{t-t06}, in the presence of a highly clumped medium. In light of this discussion and the results presented above the answer must be yes, but in these inner regions, we are still well within the outflow's sonic point (equivalent in this case to the outer shock in the \citeauthor{t-t06}~clumpy H\two\ region models) or energy injection zone \citep{shopbell98}, and at this point the gas has yet to develop into an organised flow capable of reaching into the galactic halo. We therefore interpret the broad line as arising from gas that is mixing into the high sound-speed, hot gas outflow thereby mass-loading the wind.

\begin{figure*}
\includegraphics[width=12cm]{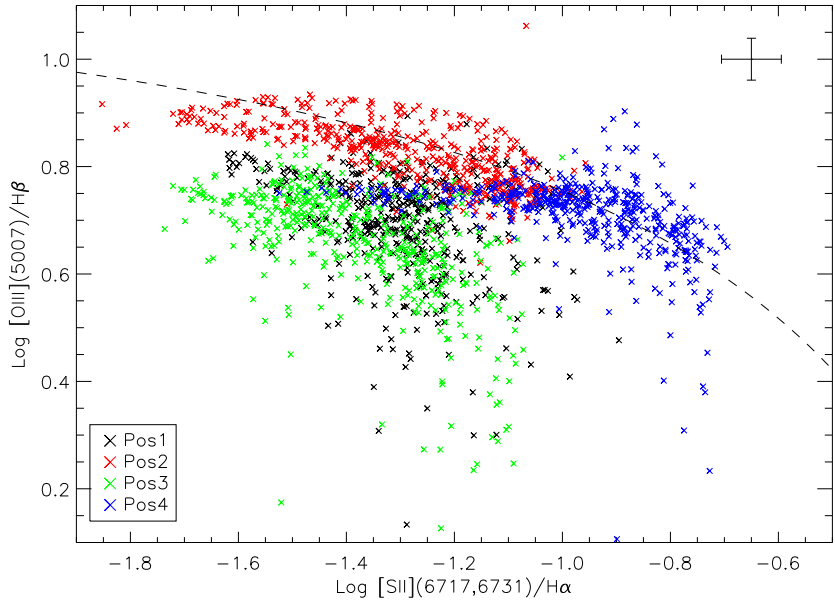}
\caption{Dereddened C1 flux ratios of [O\three]$\lambda$5007/H$\beta$ vs.~[S\two]$\lambda$6717+$\lambda$6731/H$\alpha$ for all four IFU positions. The dashed curve represents the theoretical maximum starburst line \citep{kewley01}.}
\label{fig:line_ratios_all}
\end{figure*}

Finally we turn to Fig.~\ref{fig:line_ratios_all}, which shows the [O\three]$\lambda$5007/H$\beta$ vs.~[S\two]$\lambda$6717+$\lambda$6731/H$\alpha$ nebular diagnostic diagram for all C1 measurements in all IFU positions. On this plot, we also show the theoretical `maximum starburst line' derived by \citet{kewley01}, indicating a conservative theoretical limit for pure photoionization from the hardest starburst ionizing spectrum that can be produced. For points to exist above or to the right of this threshold, an additional contribution to the excitation from non-photoionization processes (e.g.~shocks) is required. For a specific metallicity or ionization parameter, however, the hardness of the ionization field is constrained and this limit may change, but since the model predictions are quite uncertain it is difficult to know by how much. For this reason we have employed the more robust `maximum starburst line' throughout this work, as is discussed in \citetalias{westm07a}. Fig.~\ref{fig:line_ratios_all} shows that spaxels with line ratios consistent with shocks are found only in positions 2 and 4. However, what is not immediately obvious from the maps showing the spatial location of these points (fig 11 in \citetalias{westm07a} for position 2, and Fig.~\ref{fig:4elecdens} for position 4) is that the distribution of their ratios are offset, with higher [S\two]/H$\alpha$ ratios found in position 4. Since [S\two] emission can be strongly enhanced by shocks \citep{dopita97, oey00}, this position contains the strongest evidence for shock-excited emission.

\subsection{The evacuated cavity} \label{sect:cavity}
In our discussions of the individual fields, we found that the low surface-brightess inner cavity surrounding SSCs A and B can be identified at the edges of the FoV of positions 1, 3 and 4, so we should be able to make a rough assessment of the conditions in this cavity by combining the data from these small sections. Firstly, we find that all the associated corners or sections emit only faintly in H$\alpha$, as would be expected from inspection of Fig.~\ref{fig:GMOS_finder}. Furthermore, all these areas have stationary or redshifted velocities in the range 0--30~\kms{} with respect to $v_{\rm sys}$ (with the exception of C3 in the south-western corner of position 1, which has redshifted velocities of up to 100~\kms). This result is consistent with the evacuated nature of this region, implying that here we are seeing further into the galaxy to a region where the predominant gas motions are directed away from us.

In positions 1 and 4 on the north and west of SSC A, we observe very broad line widths in both C1 and C2 for areas associated with the cavity (up to 100~\kms{} in C1 and 250~\kms{} in C2), however the opposite is true in position 3, where we find the narrowest C1 line-widths of any field. Unfortunately, since our sample is very small and limited in spatial extent we cannot infer anything further. Additional observations are needed to better understand the gas conditions in this faint cavity.

\section{Summary} \label{sect:summary}

The small size of the GMOS IFU FoV ($50\times 35$~pc), coupled with the high spatial- and spectral-resolution of the instrument, have allowed us to examine each of the four fields observed in unprecedented detail, and relate the physical conditions of each field to the general state of the ISM in this central region. We now summarise the main points arising from our study.

\begin{itemize}
  \item We see knots and filaments of gas which appear to be located at different lines-of-sight and expanding in different directions, attesting to the highly disturbed and fragmented nature of the ISM in NGC 1569.
  \item We observe a narrow ($\sim$35--100~\kms) and broad ($\sim$100--400~\kms) component to the H$\alpha$ line across all four regions. The total radial velocity spread between the components is only $\sim$70~\kms.
  \item We also identify a third, high-velocity component in some regions, which we can confidently associate with shell expansion in positions 2 and 4, and possibly with the high velocity kinematic system identified by \citet{heckman95}.
  \item We confirm and expand on our conclusions of Westmoquette et al.\ (2007b, \citetalias{westm07a}) for the origins of the two main components. The most likely explanation of the narrow component (C1) is that it represents the general disturbed optically emitting ionized ISM, arising through a convolution of the stirring effects of the starburst and gravitational virial motions. This gives a region-wide characteristic turbulent width to C1 (referred to as FWHM$_{0}$), which is likely augmented by a varying contribution from unresolved kinematical components (expanding shells, knots of gas) along the line-of-sight. FWHM$_{0}$ varies between 30--45~\kms\ across the four fields, and is in good agreement with the corresponding integrated H\one\ velocity dispersions.
  \item Our data support the conclusion of \citetalias{westm07a} that the broad component (C2) results from the highly turbulent velocity field associated with the interaction of the hot phase of the ISM (the cluster winds) with cooler gas clumps -- particularly in position 1. In our model, the broadest emission is associated with the direct impact of the hot, fast-flowing cluster winds on the cool gas knots, setting up turbulent mixing layers \citep[e.g.][]{begelman90, slavin93}. Broad emission across the rest of the field most likely arises from material that is photo-evaporated or thermally evaporated from these layers through the action of the strong ambient radiation field pervading the central regions of NGC 1569, or mechanically ablated (stripped) by the impact of the winds.
  \item We find surprisingly little evidence for shocked line ratios. This may be because the signatures of shocks are overwhelmed by the strong photoionization in the central regions. Positions 2 and 4 show the most convincing evidence for shocks, where in position 4 the emission is associated with a large expanding shell complex extending to the east of the FoV.
\end{itemize}

\textit{HST} imaging (Fig.~\ref{fig:GMOS_finder}) clearly shows that the central regions of NGC 1569 are dominated by the effects of the cluster winds from SSCs A and B. This is exemplified by the swept-back, cometary-tail appearance of some of the remaining gas knots, and the existence of a large cavity surrounding the clusters, $\sim$200~pc in diameter.

This investigation has focussed on the small-scale characteristics of the ionized ISM projected within this central starburst zone, within which we find clear signatures of the multiphase nature of the ISM. We interpret the bright narrow emission lines as arising on photoionized surfaces of clouds, and find them to be kinematically associated with the H\one. While some indications of normal expanding shells also are detected, these are relatively isolated and too slow to be important in driving the starburst wind. The energetic nature of this region reveals itself through the ubiquitous presence of broad components of optical emission lines with velocity FWHM ranging from 100 to $\approx$400~\kms. The characteristic velocities of this material correspond to temperatures of 10$^5$ -- $>$10$^6$~K, far higher than the $\sim$$10^4$~K associated with the emission lines that we observed.

The velocity offsets between broad and narrow components are modest, $<$50~\kms, meaning that there is little evidence for significant ordered gas flows within the central zone of the NGC 1569 starburst, where energy is being injected into the outflow. The flow dominated wind in NGC 1569 must therefore form well beyond the region containing the massive star clusters. We interpret the combination of single peaked broad optical emission line components and the small net radial velocity offsets as arising from gas that is mixing into a high sound-speed, hot gas phase, whose presence dominates X-ray images of NGC 1569. The lack of bulk motions implies that this region is in a quasi-static state, as is expected for the inner regions of galactic winds \citep[e.g.][]{leer80, suchkov96} which pass through a sonic point. In this model the sonic point lies beyond the regions we observed, at distances of $>$100--200~pc from the starburst. In a forthcoming contribution (Westmoquette et al. in prep.; Paper III) we will examine these conclusions in the context of the large-scale galactic outflow using deep H$\alpha$ imaging and IFU observations.

Finally, we consider the wider implications of our findings. Underlying broad components to H$\alpha$ emission lines have been reported in a number of studies of giant extragalactic H\two\ regions and violent starburst regions in galaxies \citep[e.g.][]{chuken94, mendez97, homeier99, sidoli06}. Recently, we have added M82 to this list; from \textit{HST}/STIS spectroscopy we find a ubiquitous broad H$\alpha$ component throughout the starburst core \nocite{westm07c}Westmoquette et al.\ (2007b). The properties of these broad components resemble those we have found in NGC 1569 in that, compared to the narrow component, they have similar ionization conditions, radial velocities, and are spatially extended over the star-forming knots which contain young massive clusters \citep[e.g.][]{mendez97}. It therefore seems likely that in these systems too, the broad emission line component is associated with the impact of cluster winds on cool gas knots. This suggests that the broad component may well serve as a powerful diagnostic of the galactic wind phenomenon since it probably traces both mass loading of wind material, as shown here, and mass entrainment. The relationships between the properties of the broad component, the starburst regions, and the galactic wind flow merits further observational study.

\section*{Acknowledgements}
We thank the referee for his comments that have improved the clarity of the paper.


MSW thanks the Instituto de Astrof\'isca de Canarias (IAC) for their warm hospitality and financial support during the writing of this paper. KME acknowledges the support from the Euro3D Research Training Network, grant no.\ HORN-CT2002-00305.
The Gemini Observatory is operated by the Association of Universities for Research in Astronomy, Inc., under a cooperative agreement with the NSF on behalf of the Gemini partnership: the National Science Foundation (United States), the Particle Physics and Astronomy Research Council (United Kingdom), the National Research Council (Canada), CONICYT (Chile), the Australian Research Council (Australia), CNPq (Brazil) and CONICET (Argentina).

\bibliographystyle{mn2e}
\bibliography{/Users/msw/Documents/work/Thesis/thesis/references}
\bsp
\label{lastpage}

\end{document}